\documentclass[a4paper,fleqn]{cas-sc}
\usepackage{diagbox}
\usepackage[authoryear,longnamesfirst]{natbib}
\usepackage{wrapfig}
\usepackage[ruled,vlined]{algorithm2e}
\usepackage{algpseudocode} 
\usepackage{graphicx}
\graphicspath{ {Figures/} }
\usepackage{caption}
\usepackage{subcaption}
\usepackage{float}
\def\tsc#1{\csdef{#1}{\textsc{\lowercase{#1}}\xspace}}
\tsc{WGM}
\tsc{QE}
\tsc{EP}
\tsc{PMS}
\tsc{BEC}
\tsc{DE}

\begin{document}
\let\WriteBookmarks\relax
\def\floatpagepagefraction{1}
\def\textpagefraction{.001}
\shorttitle{}


\title [mode = title]{Quantitative Analysis of IITs' Research Growth and SDG Contributions}                      
\author[1,2]{Kiran Sharma}\corref{cor2}
\ead{kiran.sharma@bmu.edu.in}

\author[1]{Akshat Nagori}
\author[1]{Manya}
\author[1]{Mehul Dubey}
\author[3]{Parul Khurana}\corref{cor1}
\ead{parul.khurana@lpu.co.in}


\address{School of Engineering \& Technology, BML Munjal University, Gurugram, Haryana-122413, India }
\address{Center for Advanced Data and Computational Science, BML Munjal University, Gurugram, Haryana-122413, India }
\address{School of Computer Applications, Lovely Professional University, Phagwara, Punjab-144411, India }
\cortext[cor1]{Corresponding author 1}
\cortext[cor2]{Corresponding author 2}

\begin{abstract}
The Indian Institutes of Technology (IITs) are vital to India’s research ecosystem, advancing technology and engineering for industrial and societal benefits. This study reviews the research performance of top IITs—Bombay, Delhi, Madras, Kharagpur, and Kanpur based on Scopus-indexed publications (1952–2024). Research output has grown exponentially, supported by increased funding and collaborations. IIT-Kanpur excels in research impact, while IIT-Bombay and IIT-Madras are highly productive but show slightly lower per-paper impact. Internationally, IITs collaborate robustly with the USA, Germany, and the UK, alongside Asian nations like Japan and South Korea, with IIT-Madras leading inter-IIT partnerships. Research priorities align with SDG 3 (Health), SDG 7 (Clean Energy), and SDG 11 (Sustainable Cities). Despite strengths in fields like energy, fluid dynamics, and materials science, challenges persist, including limited collaboration with newer IITs and gaps in emerging fields. Strengthening specialization and partnerships is crucial for addressing global challenges and advancing sustainable development.

\end{abstract}


\maketitle
\section{Introduction}

India, as an emerging global power, has recognized the importance of research and is making concerted efforts to strengthen its academic and scientific ecosystem. Government initiatives such as increased funding for higher education, the establishment of research-centric policies, and global collaborations reflect this commitment~\citep{raaj2024education}.

Research not only advances scientific understanding but also delivers tangible benefits to society, such as sustainable energy solutions, improved healthcare, and technological innovations. In addition, impactful research promotes economic development by creating industries, generating employment and addressing pressing social issues such as poverty, inequality, and climate change~\citep{saini2020drives}. The ability to translate academic insights into real-world applications directly improves the well-being of communities and strengthens the global standing of the nation~\citep{jalal2020research}.

India’s premier institutes, the Indian Institutes of Technology (IIT), have been consistently at the forefront of cutting-edge research. These institutions are known for their contributions to diverse fields such as artificial intelligence, clean energy, biotechnology, and advanced manufacturing~\citep{Ghosh2021Bibliometric}.  IITs have not only produced groundbreaking research, but have also cultivated an innovation ecosystem that has led to startups, patents, and technology transfer~\citep{Nair2015Innovation}. In Indian education, IITs symbolize excellence in education and research, often being considered centers of intellectual and technological prowess.

Research conducted by IITs creates a ripple effect on innovation, driving comprehensive growth by seamlessly connecting academic exploration with industrial applications. From designing affordable healthcare solutions for underserved communities to pioneering green technologies that address environmental issues, IIT research exemplifies how academia can contribute significantly to society's welfare and national development~\citep{Chatterjee2014Trends}. As India aspires to become a global leader in science and technology, the role of IITs in the advancement of impactful research and the promotion of innovation remains critical, contributing not only to the country's economic growth, but also to its social transformation~\citep{Cheah2016Framework}.

\section{Literature review}

Indian Institutes of Technology (IITs) have consistently been recognized as leaders in innovation, research, and academic excellence. The study by \cite{Chaurasia2014Research} provides an insightful evaluation of the productivity of research and the impact of IIT Delhi over a decade. The study concludes that IIT Delhi’s research during this period demonstrated significant growth, with increasing contributions to science and technology. Collaboration and interdisciplinary research emerged as key strengths that enhanced the institution's academic reputation, with an emphasis on international partnerships to boost research impact.

The bibliometric study by \cite{awasthi2020highly} highlighted the key role of IITs in India's scientific growth and global research, emphasizing quality and highly cited and influential publications. The study by \cite{Siddaiah2016Contribution} analyzes the research output and the impact of the citation of eight newly established IITs during 2010–2014. Although relatively new, these IITs demonstrated an increase in research productivity, focusing on engineering, physics, and materials science. Collaborative efforts, especially with established IITs and international partners, greatly improved the impact of citations and the visibility of research. The study highlights the competitive quality of research from these institutions and recommends strengthening research infrastructure and collaborations to further their contributions to the scientific landscape of India. 

\cite{Pradhan2018Bibliometric} conducted a bibliometric analysis of IIT research publications indexed in Scopus, highlighting trends in productivity, citation impact, and collaborative patterns. The study reveals a significant growth in the publication output, with engineering and technology leading the research domains. International collaborations and high-impact publications underscore the global relevance of IIT research. \cite{Ramesh2017Scientometrics} conducted a scientometric analysis of engineering research at IIT Madras and IIT Bombay from 2006 to 2015. The study also reports a constant growth in research output, higher shares of engineering disciplines, and a high citation impact. Both institutes had wide international collaborations, which gave high relevance and visibility to their research output. The authors underline the strategic partnership and funding as main drivers for sustaining research excellence.

Unique contributions of every individual IIT in research must be studied in depth, while understanding their strengths and weaknesses, their standing in terms of global competitiveness, and their national priority linkages. It also fosters accountability and informs policy decisions related to funding, collaborations, and resource allocation. IIT Bombay has been a leader in database systems and data management research since the 1980s, contributing significantly to information retrieval and data mining~\citep{Chakrabarti2013}. The IIT Bombay Developmental Informatics Lab focuses on leveraging ICT (Information and Communication Technology) to improve access to information in rural India, addressing critical needs such as agriculture and tribal education~\citep{Bahuman2007}. ~\cite{Hasan2015} provides a scientometric analysis of the research output of the leading IITs over five years. The research output of the five best performing IITs accounted for 9. 32\% of the total Indian research output, with a maximum of 22 27\% articles indexed in 2013. \cite{Das2022Trendline} has studied that IIT Hyderabad (26\%), with an increase in open-access publications, is the top research institute in India for open-access journals. \cite{Ghosh2021Bibliometric} conducted a bibliometric analysis of research productivity in physics, chemistry, and mathematics at IIT Kharagpur from 2001 to 2020. The study highlights significant growth in publication output, with physics leading in productivity and impact of citations. Collaborative research, both national and international, played a key role in enhancing research visibility and impact. The findings underscore the importance of interdisciplinary collaboration and sustained support for research in fundamental sciences. \cite{Bhui2018Publications} conducted a bibliometric study of publications by faculty members in the Department of Humanities and Social Sciences of IIT Kharagpur. The analysis highlights the research output trends, with a focus on journal articles and conference papers. The study reveals increasing interdisciplinary research and moderate citation impact, emphasizing the growing contribution of HSS to IIT Kharagpur's academic landscape. 

Similarly, other higher education institutes such as IISER, NIT,etc. were also analyzed in terms of their performance. \cite{Solanki2016Research} evaluated the research competitiveness of IISERs through publications and citations, highlighting their rapid growth and notable contributions in chemistry, physics, and biology despite their status as relatively young institutions. \citep{Bala2013Research} analyze the research performance of the National Institutes of Technology (NITs) of India from 2001 to 2010 using bibliometric methods. The study underscores the consistent growth in research output, with engineering and technology leading the publication domains. Collaborative efforts, both nationally and internationally, have markedly improved the impact of their work.

IITs collectively exhibit significant research contributions in Computer Science, with robust bibliometric indicators such as citation rates and h-index scores, reflecting their global impact~\citep{Singh2019, Arif2015}. \cite{Krishna2009, Chandra2010} examined the role of IITs in fostering university-industry collaborations and cultivating an entrepreneurial culture, thus improving India's innovation ecosystem.\citep{Prathap2018} compared the impact of IIT research with leading global institutions in engineering. \cite{boshoff2020conceptualizing} 
analyzed the social impact of research by presenting it through results, outcomes, and larger societal benefits. They highlight that such research drives progress in public health, education, and economic development while promoting behavioral changes and improving quality of life.

In line with the 17 SDGs of the United Nations, the research output of IITs has contributed substantially to the tackling of critical global challenges, including healthcare, clean energy, sustainable cities and climate action. These institutions play a central role in the advancement of knowledge, the driving force of technological innovation, and the implementation of sustainable solutions to meet the demands of national and global sustainability agendas~\cite{priyadarshini2020piecemeal, singh2022research}.
\subsection{Research gap}

Although numerous studies have examined the research productivity and impact of IITs, key aspects are still not adequately explored. Previous literature has predominantly focused on overall productivity, citation metrics, and thematic strengths, often overlooking nuanced areas such as the role of interdisciplinary collaborations, the impact of newer IITs, and the alignment of research with emerging global challenges. Furthermore, there is limited exploration of the contribution of IITs to the achievement of sustainable Development Goals (SDGs).

Unlike previous studies that focus on short-term productivity, this study examines seven decades of research output, capturing historical growth trajectories and transformative periods. Our findings also highlight the gaps in collaborative networks, showing limited partnerships between older and newer IITs. By aligning IIT research contributions with SDGs, the study provides information on their role in addressing global challenges along with their international partners. These contributions offer a deeper understanding of IIT research while addressing critical literature gaps, guiding strategies for enhanced collaboration, emerging trends, and social impact.
\subsection{Research objectives}
\begin{itemize}
    \item Analyze the growth trajectory of research publications in the five holder IITs (Bombay, Delhi, Madras, Kharagpur and Kanpur) from 1952 to 2024.
    \item To examine productivity trends and measure citation impacts of publications for each IIT.
    \item Identify patterns of inter-IIT and international collaborations and their impact on research outcomes.
    \item Uncover unique research themes and interdisciplinary approaches among the IITs.
    \item Evaluate the contribution of IIT research to the achievement of the SDGs.
\end{itemize}
\section{Methodology}

\begin{figure}[!h]
    \centering
\includegraphics[width=0.4\linewidth]{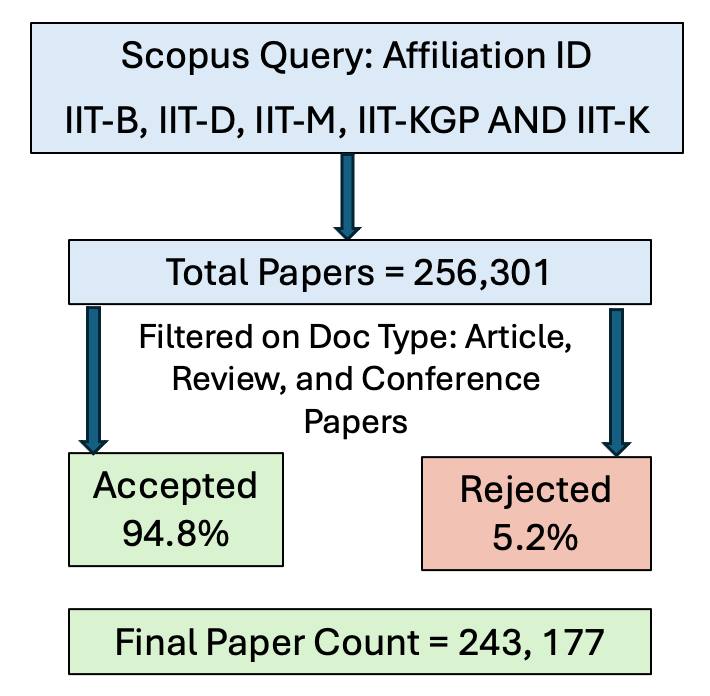} 
\caption{Data downloading and filtering flowchart.}
\label{fig:flowchart}
\end{figure}

Data for the Indian Institutes of Technology (IIT) collaboration analysis were sourced from Scopus, a leading database for peer-reviewed academic literature. The query was constructed using unique Scopus affiliation IDs for IIT-Bombay (IIT-B), IIT-Delhi (IIT-D), IIT-Madras (IIT-M), IIT-Kharagpur (IIT-KGP) and IIT-Kanpur (IIT-K) to specifically capture publications authored by researchers affiliated with these institutions. The search was further refined to include only three types of documents, articles, reviews, and conference papers, as these represent the core academic outputs that reflect substantial research contributions. The query was executed in September 2024, ensuring that the dataset contained the most recent publications available at the time. The resulting dataset included metadata such as publication titles, authors, affiliations, collaboration details, publication years, document types, etc. Figure~\ref{fig:flowchart} demonstrates the data downloading and filtering process. 

Table~\ref{table:dataDesc} provides a data description of research publications from five IITs. Details the year of establishment, affiliation ID, the total number of papers published, the filtered papers, and the percentage of papers accepted and discarded. IIT-KGP (1951), the oldest institute among the five listed IITs, and IIT-D (1961) is the youngest among five.
\begin{table}[!h]
\centering
\caption{Data description.}
\begin{tabular}{|l|c|c|c|c|c|c|c|}
\hline
\textbf{Institute Name}                  & \textbf{Abbv.} & \textbf{\begin{tabular}[c]{@{}c@{}}Estb \\ Year\end{tabular}} & \textbf{\begin{tabular}[c]{@{}c@{}}Affiliation \\ ID\end{tabular}} & \textbf{\begin{tabular}[c]{@{}c@{}}Total\\ Papers\end{tabular}} & \textbf{\begin{tabular}[c]{@{}c@{}}Filtered\\ Papers\end{tabular}} & \textbf{\begin{tabular}[c]{@{}c@{}}\% \\ Accepted\end{tabular}} & \textbf{\begin{tabular}[c]{@{}c@{}}\%\\ Discarded\end{tabular}} \\ \hline
Indian Institute of Technology-Bombay    & IIT-B          & \textbf{1958}                                                 & 60014153                                                           & 50257                                                           & 47785                                                              & 95.08                                                           & 4.92                                                            \\ \hline
Indian Institute of Technology-Delhi     & IIT-D          & \textbf{1961}                                                 & 60032730                                                           & 56775                                                           & 53548                                                              & 94.32                                                           & 5.68                                                            \\ \hline
Indian Institute of Technology-Madras    & IIT-M          & \textbf{1959}                                                 & 60025757                                                           & 52386                                                           & 50027                                                              & 95.50                                                           & 4.50                                                            \\ \hline
Indian Institute of Technology-Kharagpur & IIT-KGP        & \textbf{1951}                                                 & 60004750                                                           & 56180                                                           & 53232                                                              & 94.75                                                           & 5.25                                                            \\ \hline
Indian Institute of Technology-Kanpur    & IIT-K          & \textbf{1959}                                                 & 60021988                                                           & 40703                                                           & 38585                                                              & 94.80                                                           & 5.20                                                            \\ \hline
\end{tabular}
\label{table:dataDesc}
\end{table}
 Publishing in open-access journals often requires article processing charges (APCs), which may not be fully covered by institutional funding, leading to lower open-access publication percentages. Table~\ref{table:openAccess} shows data on the number of research papers published by various IITs and their accessibility. Across all institutes, a smaller percentage of papers (arond 10-12\%) are open access compared to non-open access, suggesting that most research output is not freely accessible. For IIT-B, open-access papers have a higher citation rate than nonopen-access papers, suggesting that they might be reaching a broader audience or have higher visibility. In general, non-open-access papers have slightly higher citation rates than open-access papers, possibly because they are published in journals with established academic readerships. 

\begin{table}[!h]
\centering
\caption{Distribution of open access and no open access publications of top 5 IITs.}
\begin{tabular}{|l|c|cccc|cccc|}
\hline
\multirow{2}{*}{\textbf{Institute}} & \multirow{2}{*}{\textbf{\begin{tabular}[c]{@{}c@{}}Total \\ Papers\end{tabular}}} & \multicolumn{4}{c|}{\textbf{Open Access}} & \multicolumn{4}{c|}{\textbf{No Open Access}}   \\ \cline{3-10} 
 &  & \multicolumn{1}{c|}{\textbf{\begin{tabular}[c]{@{}c@{}}Paper \\ Count\end{tabular}}} & \multicolumn{1}{c|}{\textbf{\begin{tabular}[c]{@{}c@{}}\%Paper\\ Count\end{tabular}}} & \multicolumn{1}{c|}{\textbf{\begin{tabular}[c]{@{}c@{}}Total \\ Citations\end{tabular}}} & \textbf{\begin{tabular}[c]{@{}c@{}}Paper-\\ Citation \\ Ratio\end{tabular}} & \multicolumn{1}{c|}{\textbf{\begin{tabular}[c]{@{}c@{}}Paper \\ Count\end{tabular}}} & \multicolumn{1}{c|}{\textbf{\begin{tabular}[c]{@{}c@{}}\%\\ Paper\end{tabular}}} & \multicolumn{1}{c|}{\textbf{\begin{tabular}[c]{@{}c@{}}Total \\ Citations\end{tabular}}} & \textbf{\begin{tabular}[c]{@{}c@{}}Paper-\\ Citation \\ Ratio\end{tabular}} \\ \hline
\textbf{IIT-B}                      & 47785     & \multicolumn{1}{c|}{6200}   & \multicolumn{1}{c|}{12.98}  & \multicolumn{1}{c|}{133807}                                                              & 21.58                                                                       & \multicolumn{1}{c|}{41585} & \multicolumn{1}{c|}{87.02}   & \multicolumn{1}{c|}{732458}                                                              & 17.61  \\ \hline
\textbf{IIT-D}                      & 53548                                                                             & \multicolumn{1}{c|}{5898}         & \multicolumn{1}{c|}{11.01}& \multicolumn{1}{c|}{115373}& 19.57   & \multicolumn{1}{c|}{47650}   & \multicolumn{1}{c|}{88.99}                                                       & \multicolumn{1}{c|}{947281}  & 19.88                                                                       \\ \hline
\textbf{IIT-M}                      & 50027                                                                             & \multicolumn{1}{c|}{5926}                                                            & \multicolumn{1}{c|}{11.85}                                                       & \multicolumn{1}{c|}{98505}   & 16.63                                                                       & \multicolumn{1}{c|}{44101}                                                           & \multicolumn{1}{c|}{88.15}                                                       & \multicolumn{1}{c|}{756223}                                                              & 17.14  \\ \hline
\textbf{IIT-KGP}                    & 53232                                                                             & \multicolumn{1}{c|}{5499}                                                            & \multicolumn{1}{c|}{10.33} & \multicolumn{1}{c|}{109076} & 19.83    & \multicolumn{1}{c|}{47733}   & \multicolumn{1}{c|}{89.67}                                                       & \multicolumn{1}{c|}{973355}  & 20.39    \\ \hline \textbf{IIT-K}                      & 38585        & \multicolumn{1}{c|}{4035} & \multicolumn{1}{c|}{10.46}  & \multicolumn{1}{c|}{70731}        & 17.53  & \multicolumn{1}{c|}{34550} & \multicolumn{1}{c|}{89.54}  & \multicolumn{1}{c|}{741105}     & 21.45    \\ \hline
\end{tabular}
\label{table:openAccess}
\end{table}
Table~\ref{table:docType} presents data on the distribution of academic papers from five IITs. The data is divided into three categories: Articles, Reviews, and Conference Papers. Articles form the majority of publications in all institutes, contributing approximately 70–77\% of the total articles. Reviews constitute a much smaller fraction, around 2–3\% of the total papers. Conference Papers make up a significant portion, about 20–27\%, depending on the institute. IIT-D leads with 55,348 total published papers, and IIT-K has the lowest total, at 38,585. IIT-KGP and IIT-M are close in total papers, with IIT-KGP (53,232) slightly ahead of IIT-M (50,027).
IIT-B ranks fourth with 47,785 total papers. While all IITs show strong research output, IIT-KGP excels in terms of overall research impact (citations), particularly in articles and reviews. IIT-D and IIT-M show balanced contributions across all categories, while IIT-B demonstrates a strong focus on impactful conference papers. IIT-K, although it has published fewer papers, maintains a steady presence and has potential for growth in impact.

In addition, IIT-KGP has the highest overall citation-to-paper ratio (22.04), driven by its strong performance in articles and reviews. The citation-to-paper ratio provides an indicator of the average impact of each paper, measuring how frequently each paper is cited. The citation-to-paper ratio highlights IIT-KGP as the leader in research impact, particularly in articles and reviews, while IIT-D excel in reviews and conference papers.


\begin{table}[!h]
\centering
\caption{Distribution of papers as per document type:article, review, and conference paper.}
\begin{tabular}{|l|c|ccc|ccc|ccc|}
\hline
\multicolumn{1}{|c|}{\multirow{2}{*}{\textbf{Institute}}} & \multirow{2}{*}{\textbf{\begin{tabular}[c]{@{}c@{}}Total \\ Papers\end{tabular}}} & \multicolumn{3}{c|}{\textbf{Article}}                                                                                                                                                                                                        & \multicolumn{3}{c|}{\textbf{Review}} & \multicolumn{3}{c|}{\textbf{Conference Paper}}\\ \cline{3-11} 
\multicolumn{1}{|c|}{}                                    &                                                                                   & \multicolumn{1}{c|}{\textbf{\begin{tabular}[c]{@{}c@{}}Paper\\ Count\end{tabular}}} & \multicolumn{1}{c|}{\textbf{\begin{tabular}[c]{@{}c@{}}\%Paper\\ Count\end{tabular}}} & \textbf{\begin{tabular}[c]{@{}c@{}}Total \\ Citations\end{tabular}} & \multicolumn{1}{c|}{\textbf{\begin{tabular}[c]{@{}c@{}}Paper\\ Count\end{tabular}}} & \multicolumn{1}{c|}{\textbf{\begin{tabular}[c]{@{}c@{}}\%\\ Paper\end{tabular}}} & \textbf{\begin{tabular}[c]{@{}c@{}}Total \\ Citations\end{tabular}} & \multicolumn{1}{c|}{\textbf{\begin{tabular}[c]{@{}c@{}}Paper\\ Count\end{tabular}}} & \multicolumn{1}{c|}{\textbf{\begin{tabular}[c]{@{}c@{}}\%\\ Paper\end{tabular}}} & \textbf{\begin{tabular}[c]{@{}c@{}}Total \\ Citations\end{tabular}} \\ \hline
\textbf{IIT-B}                                            & 47785                                                                             & \multicolumn{1}{c|}{33474}                                                          & \multicolumn{1}{c|}{70.05}                                                       & 718661                                                              & \multicolumn{1}{c|}{1224}                                                           & \multicolumn{1}{c|}{2.56}                                                        & 71349                                                               & \multicolumn{1}{c|}{13087}                                                          & \multicolumn{1}{c|}{27.39}                                                       & 76255                                                               \\ \hline
\textbf{IIT-D}                                            & 53548                                                                             & \multicolumn{1}{c|}{39316}                                                          & \multicolumn{1}{c|}{73.42}                                                       & 874685                                                              & \multicolumn{1}{c|}{1748}                                                           & \multicolumn{1}{c|}{3.27}                                                        & 107347                                                              & \multicolumn{1}{c|}{12484}                                                          & \multicolumn{1}{c|}{23.31}                                                       & 80622                                                               \\ \hline
\textbf{IIT-M}                                            & 50027                                                                             & \multicolumn{1}{c|}{36904}                                                          & \multicolumn{1}{c|}{73.77}                                                       & 735398                                                              & \multicolumn{1}{c|}{1102}                                                           & \multicolumn{1}{c|}{2.2}                                                         & 61304                                                               & \multicolumn{1}{c|}{12021}                                                          & \multicolumn{1}{c|}{24.03}                                                       & 58026                                                               \\ \hline
\textbf{IIT-KGP}                                          & 53232                                                                             & \multicolumn{1}{c|}{41017}                                                          & \multicolumn{1}{c|}{77.05}                                                       & 930932                                                              & \multicolumn{1}{c|}{1452}                                                           & \multicolumn{1}{c|}{2.73}                                                        & 89064                                                               & \multicolumn{1}{c|}{10763}                                                          & \multicolumn{1}{c|}{20.22}                                                       & 62435                                                               \\ \hline
\textbf{IIT-K}                                            & 38585                                                                             & \multicolumn{1}{c|}{29353}                                                          & \multicolumn{1}{c|}{76.07}                                                       & 696738                                                              & \multicolumn{1}{c|}{881}                                                            & \multicolumn{1}{c|}{2.28}                                                        & 58597                                                               & \multicolumn{1}{c|}{8351}                                                           & \multicolumn{1}{c|}{21.65}                                                       & 56141 \\ \hline
\end{tabular}
\label{table:docType}
\end{table}


\section{Results and Discussion}
\subsection{Publication trend analysis}
\begin{figure}[!h]
    \centering
\includegraphics[width=0.85\linewidth]{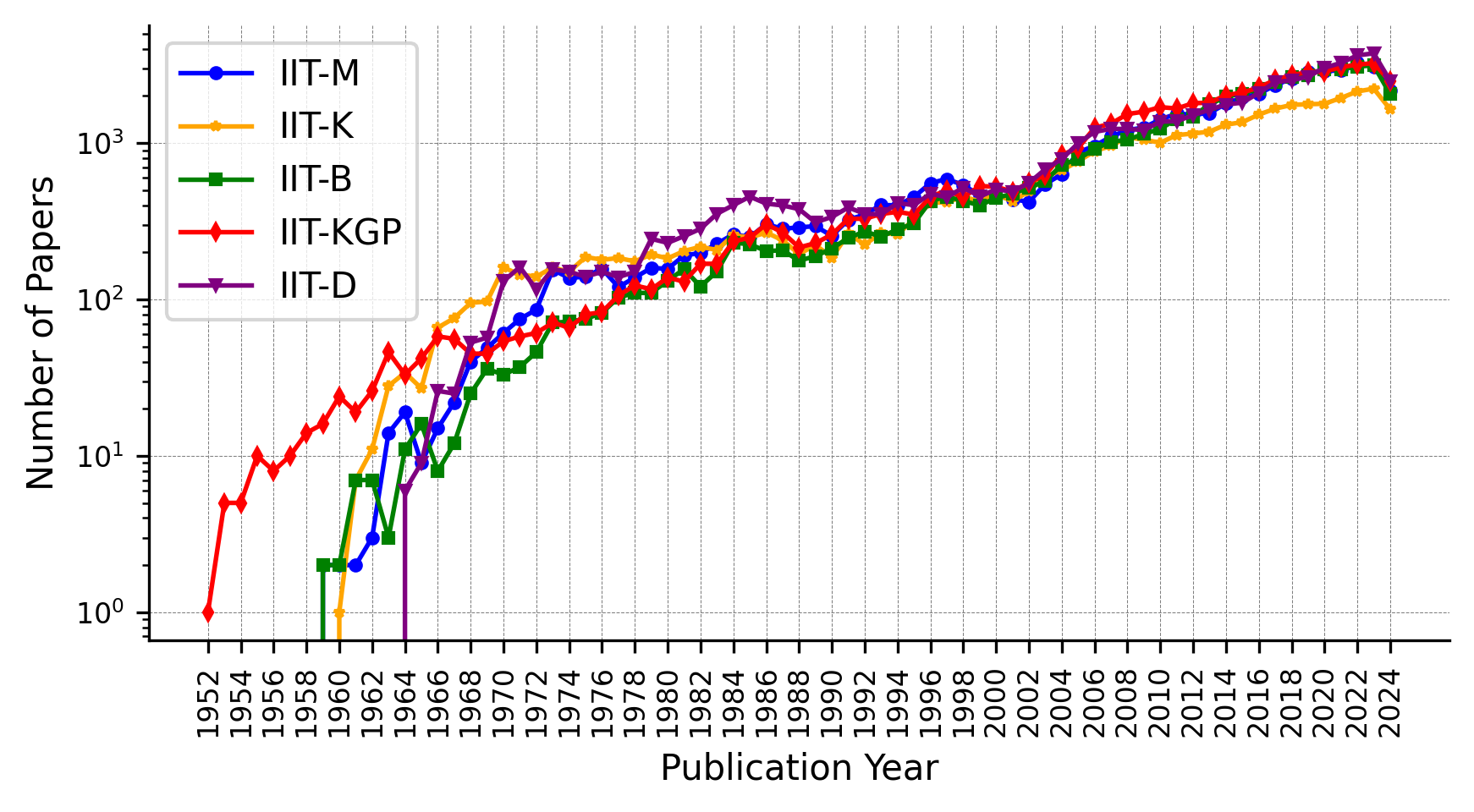} 
\caption{Year-wise publication trends of top 5 ITT's: Madras, Delhi, Bombay, Kharagpur,and Kanpur.}
\label{fig:yearPub}
\end{figure}

Figure~\ref{fig:yearPub} shows the long-term growth trajectory of research outputs from major IITs. This demonstartes the yearly trend in the number of research papers published by five IITs from 1952 to 2024. IIT-KGP led in research output initially, as it was established earlier (in 1951). Other IITs, such as IIT-B and IIT-D, show a slower start in research publications, likely due to being established later. All IITs show steady growth in research output, moving from dozens to hundreds of papers per year. IIT-KGP maintained a leading position during this period, but other IITs began to close the gap. By the late 1980s and early 1990s, IIT-B and IIT-M had established themselves as competitive research institutions, reaching publication levels similar to IIT-KGP. This period marks a consistent increase across all IITs, reflecting their expanding research focus and resources. There is a notable increase in the number of publications across all IITs, especially from 2000 onward, indicating a surge in research activity. By 2008, the publication output for each IIT reaches more than 1000 papers per year, reflecting enhanced research funding, resources, and collaborative projects. The logarithmic scale emphasizes how each IIT has grown from publishing a handful of papers annually to publishing thousands, marking a substantial rise in their global research contributions.

\begin{figure}[!h]
    \centering
\includegraphics[width=0.85\linewidth]{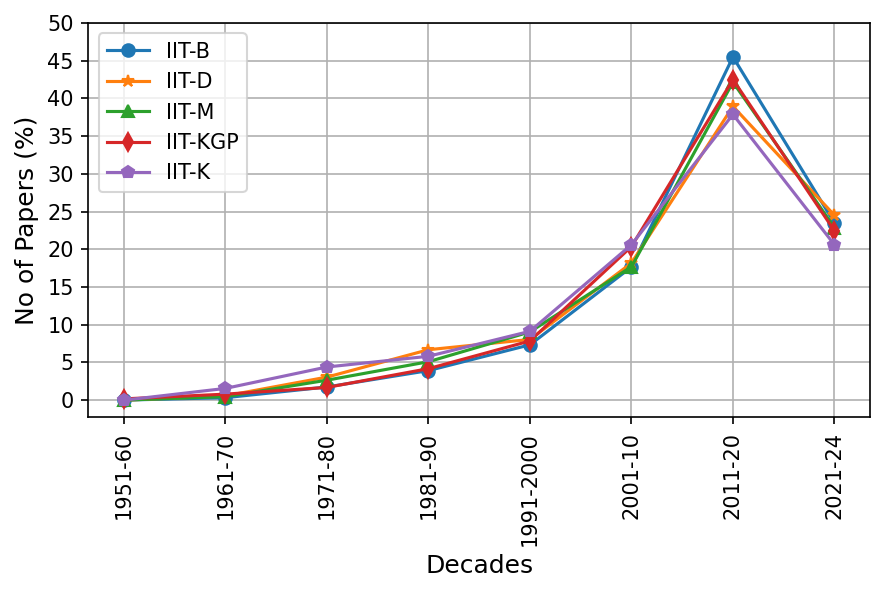} 
\caption{Decade-wise publication trends of top 5 ITT's.}
\label{fig:decadeTrend}
\end{figure}

Figure~\ref{fig:decadeTrend} gives a comparative view of research growth in major IITs over the decades. This shows the trend in the percentage of research papers published by five IITs from 1951 to 2024. From 1951 to around 2000, the number of papers published by each IIT remained relatively low and showed only a slight increase. This period reflects the early growth stage of research publications in these institutions. In the 2001–2010 decade, the publication percentages for each IIT began to increase more noticeably, indicating a growth in research output. This decade shows a sharp increase in the number of published papers, with the five IITs reaching their highest publication percentages around 2011–2020. IIT-B, in particular, reached the highest percentage of around 45\%, leading the group.

\begin{figure}[!h]
    \centering
\includegraphics[width=0.85\linewidth]{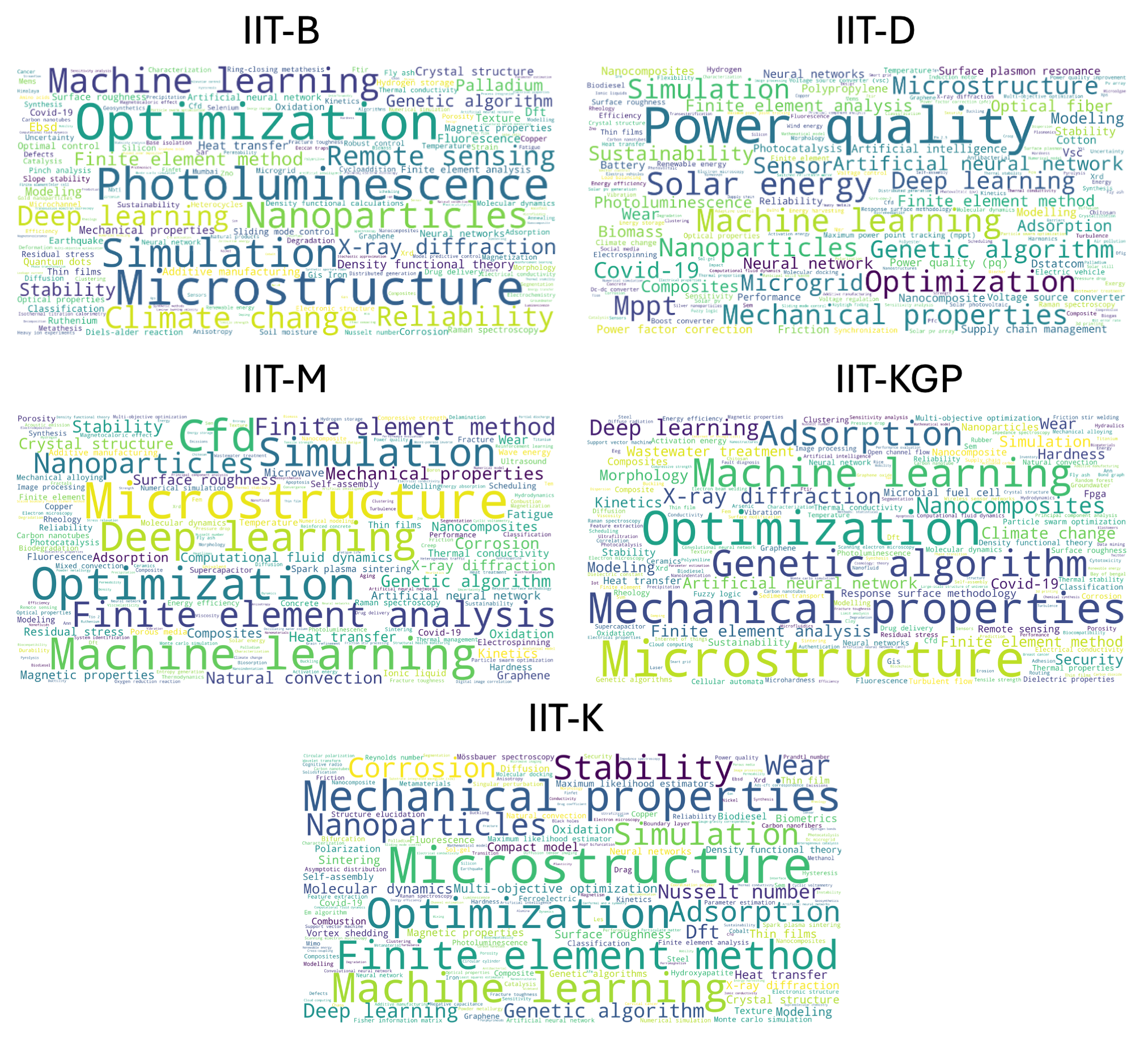} 
\caption{Thematic analysis of research produced by all IITs.}
\label{fig:wordCloud}
\end{figure}

In addition, Table~\ref{table:decadePub} provides a decade-wise distribution of the total number of publications and total citations for the five IITs in the top five from 1951 to 2024. Each row represents a decade, showing both the number of published papers (Total Papers) and the cumulative number of times these papers were cited (Total Citations). All IITs show a marked increase in both publications and citations over time, with particularly rapid growth in the 21st century. IIT-B and IIT-M have become leaders in both publications and citations, especially in the last two decades. However, IITs such as IIT-K and IIT-D, despite lower publication counts, have higher citation-to-publication ratios, indicating impactful and highly regarded research. Although IIT-KGP initially led both in publications and citations, IIT-B, IIT-M, and IIT-D have caught up and, in some cases, surpassed IIT-KGP in recent decades, highlighting the dynamic nature of research productivity across IITs. This suggests that, although IIT-K publishes fewer papers compared to some other IITs, its research is highly influential. IIT-KGP follows closely with a ratio of 20.33, showing strong influence and a significant number of citations per paper. IIT-D has a ratio of 19.85, indicating impactful research, slightly lower than IIT-K and IIT-KGP. IIT-B and IIT-M have relatively lower ratios, at 18.13 and 17.09 respectively. Although they have high total citations and publication numbers, their average citations per paper are slightly lower than those of IIT-K, IIT-KGP, and IIT-D.
\begin{table}[!h]
\centering
\caption{Decade-wise distribution of number of publications and total citations for five IIT's.}
\begin{tabular}{|l|cc|cc|cc|cc|cc|}
\hline
\multirow{2}{*}{\textbf{Decade}} & \multicolumn{2}{c|}{\textbf{IIT-B}}& \multicolumn{2}{c|}{\textbf{IIT-D}} & \multicolumn{2}{c|}{\textbf{IIT-M}}& \multicolumn{2}{c|}{\textbf{IIT-KGP}}  & \multicolumn{2}{c|}{\textbf{IIT-K}}  \\ \cline{2-11} 
& \multicolumn{1}{c|}{\textbf{\begin{tabular}[c]{@{}c@{}}Total\\ Papers\end{tabular}}} & \textbf{\begin{tabular}[c]{@{}c@{}}Total \\ Citations\end{tabular}} & \multicolumn{1}{c|}{\textbf{\begin{tabular}[c]{@{}c@{}}Total\\ Papers\end{tabular}}} & \textbf{\begin{tabular}[c]{@{}c@{}}Total \\ Citations\end{tabular}} & \multicolumn{1}{c|}{\textbf{\begin{tabular}[c]{@{}c@{}}Total\\ Papers\end{tabular}}} & \textbf{\begin{tabular}[c]{@{}c@{}}Total \\ Citations\end{tabular}} & \multicolumn{1}{c|}{\textbf{\begin{tabular}[c]{@{}c@{}}Total\\ Papers\end{tabular}}} & \textbf{\begin{tabular}[c]{@{}c@{}}Total \\ Citations\end{tabular}} & \multicolumn{1}{c|}{\textbf{\begin{tabular}[c]{@{}c@{}}Total\\ Papers\end{tabular}}} & \textbf{\begin{tabular}[c]{@{}c@{}}Total \\ Citations\end{tabular}} \\ \hline
\textbf{1951-60}                 & \multicolumn{1}{c|}{4}                                                               & 126                                                                 & \multicolumn{1}{c|}{0}                                                               & 0                                                                   & \multicolumn{1}{c|}{4}                                                               & 0                                                                   & \multicolumn{1}{c|}{93}                                                              & 915                                                                 & \multicolumn{1}{c|}{1}                                                               & 33                                                                  \\ \hline
\textbf{1961-70}                 & \multicolumn{1}{c|}{158}                                                             & 1788                                                                & \multicolumn{1}{c|}{307}                                                             & 1811                                                                & \multicolumn{1}{c|}{234}                                                             & 1509                                                                & \multicolumn{1}{c|}{424}                                                             & 3713                                                                & \multicolumn{1}{c|}{601}                                                             & 10558                                                               \\ \hline
\textbf{1971-80}                 & \multicolumn{1}{c|}{838}                                                             & 9382                                                                & \multicolumn{1}{c|}{1636}                                                            & 16961                                                               & \multicolumn{1}{c|}{1324}                                                            & 13513                                                               & \multicolumn{1}{c|}{902}                                                             & 8960                                                                & \multicolumn{1}{c|}{1696}                                                            & 31879                                                               \\ \hline
\textbf{1981-90}                 & \multicolumn{1}{c|}{1872}                                                            & 22801                                                               & \multicolumn{1}{c|}{3580}                                                            & 56129                                                               & \multicolumn{1}{c|}{2551}                                                            & 28480                                                               & \multicolumn{1}{c|}{2226}                                                            & 30658                                                               & \multicolumn{1}{c|}{2244}                                                            & 31866                                                               \\ \hline
\textbf{1991-2000}               & \multicolumn{1}{c|}{3509}                                                            & 89725                                                               & \multicolumn{1}{c|}{4309}                                                            & 94486                                                               & \multicolumn{1}{c|}{4528}                                                            & 96939                                                               & \multicolumn{1}{c|}{4177}                                                            & 85433                                                               & \multicolumn{1}{c|}{3524}                                                            & 97981                                                               \\ \hline
\textbf{2001-10}                 & \multicolumn{1}{c|}{8438}                                                            & 242567                                                              & \multicolumn{1}{c|}{9733}                                                            & 321868                                                              & \multicolumn{1}{c|}{8849}                                                            & 249262                                                              & \multicolumn{1}{c|}{10841}                                                           & 358269                                                              & \multicolumn{1}{c|}{7954}                                                            & 294720                                                              \\ \hline
\textbf{2011-20}                 & \multicolumn{1}{c|}{21740}                                                           & 438399                                                              & \multicolumn{1}{c|}{20856}                                                           & 477319                                                              & \multicolumn{1}{c|}{21089}                                                           & 403278                                                              & \multicolumn{1}{c|}{22624}                                                           & 508498                                                              & \multicolumn{1}{c|}{14616}                                                           & 299083                                                              \\ \hline
\textbf{2021-24}                 & \multicolumn{1}{c|}{11226}                                                           & 61477                                                               & \multicolumn{1}{c|}{13127}                                                           & 94080                                                               & \multicolumn{1}{c|}{11448}                                                           & 61747                                                               & \multicolumn{1}{c|}{11945}                                                           & 85985                                                               & \multicolumn{1}{c|}{7949}                                                            & 45356                                                               \\ \hline
\textbf{Total}                   & \multicolumn{1}{c|}{47785}                                                           & 866265                                                              & \multicolumn{1}{c|}{53548}                                                           & 1062654                                                             & \multicolumn{1}{c|}{50027}                                                           & 854728                                                              & \multicolumn{1}{c|}{53232}                                                           & 1082431                                                             & \multicolumn{1}{c|}{38585}                                                           & 811476                                                              \\ \hline
\end{tabular}
\label{table:decadePub}
\end{table}
Figure~\ref{fig:wordCloud} shows the word clouds from five different IITs. Each word cloud represents prominent research themes at each institution, with the size of each word indicating its relative prominence.  All IITs share interests in computational methods, optimization, and materials science, each institute has unique specializations that reflect its strengths and research priorities. IIT-B has a balanced focus on materials science (photoluminescence and microstructure) and computational methods (machine learning and optimization), with additional interest in environmental themes like climate change. IIT-D emphasizes energy (Solar Energy, Power Quality), sustainability, and materials science, along with advanced computational methods.
IIT-M shows strong interest in fluid dynamics (CFD), materials (microstructure), and computational techniques (Finite Element Method, Optimization). IIT-KGP has a diverse range of themes, but places emphasis on adsorption processes, mechanical properties, and machine learning.
IIT-K focuses on corrosion, mechanical properties, and materials science, with a significant presence of computational methods such as finite element analysis. \cite{Das2002} revealed that IIT-Delhi has been actively researching and developing low-emission hydrogen powered engines for nearly two decades, with significant advancements in performance, emission and combustion characteristics.

\subsection{Inter-IIT collaboration}

\begin{figure}[!h]
    \centering
\includegraphics[width=\linewidth]{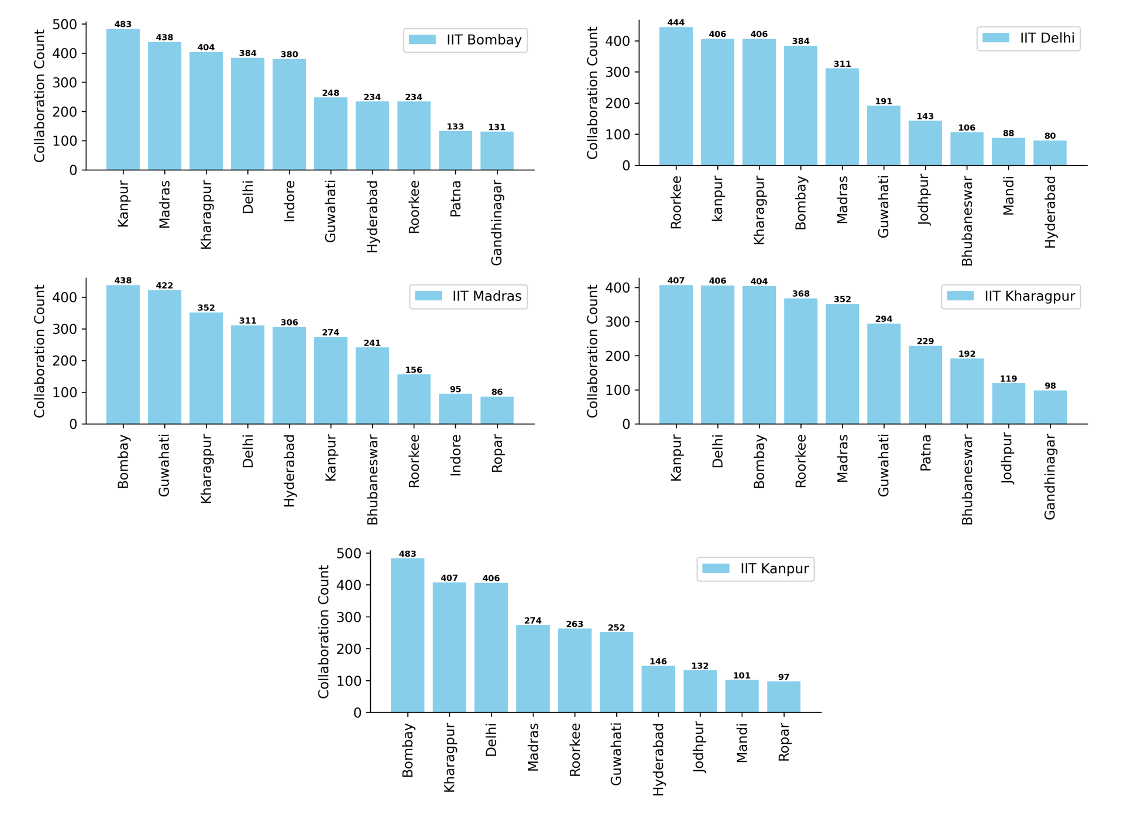} 
\caption{Collaboration of top 5 ITT's with other IITs.}
\label{fig:InterIITColab}
\end{figure}

Figure~\ref{fig:InterIITColab} displays the collaboration counts between five IITs - Bombay, Delhi, Madras, Kharagpur and Kanpur - and other IITs. Each bar plot represents the number of collaborations between a specific IIT and other IITs, sorted in descending order. Across the five IITs (Bombay, Delhi, Madras, Kharagpur, and Kanpur), collaborations are strongest with each other (the older IITs). IIT Bombay, Kanpur, and Delhi frequently appear as top collaborators in different IITs. IITs Roorkee and Guwahati also have significant collaborations, particularly with Delhi, Madras, and Kharagpur, but tend to have fewer collaborations with newer IITs.  IITs like Gandhinagar, Ropar, Jodhpur, and Mandi generally have fewer collaboration counts, indicating fewer interactions with the more established IITs. 

IIT-Bombay has its strongest collaborative relationships with IIT Kanpur, Madras, and Delhi, indicating a concentration of joint research or projects with these institutions. IIT-Delhi has its highest collaboration with IIT Roorkee, suggesting strong ties. Its collaborations are relatively well distributed among other major IITs, with lower interaction with newer or smaller IITs. IIT-Madras has strong collaborative links with IIT Bombay, Guwahati, and Kharagpur. There is less collaboration with newer IITs like Ropar and Indore. IIT-Kharagpur collaborates most frequently with IIT Kanpur, Delhi, and Bombay, showing a strong connection with the older IITs. IITs like Gandhinagar and Jodhpur have relatively fewer collaborations with Kharagpur.
IIT Kanpur has its highest collaboration with IIT Bombay, which shows a strong link. The lower collaboration counts with newer IITs, like Mandi and Ropar, indicate a preference for working with the established IITs. Overall, this analysis shows a pattern in which older and more established IITs (Bombay, Delhi, Kanpur, Kharagpur and Madras) tend to collaborate more with each other, probably due to historical relationships, larger research output and available resources. Collaborations with newer IITs are generally lower, which could be due to geographical distance, newer research programs, or differing research focuses.

\subsection{Collaboration patterns}

\begin{figure}[!h]
    \centering
\includegraphics[width=0.8\linewidth]{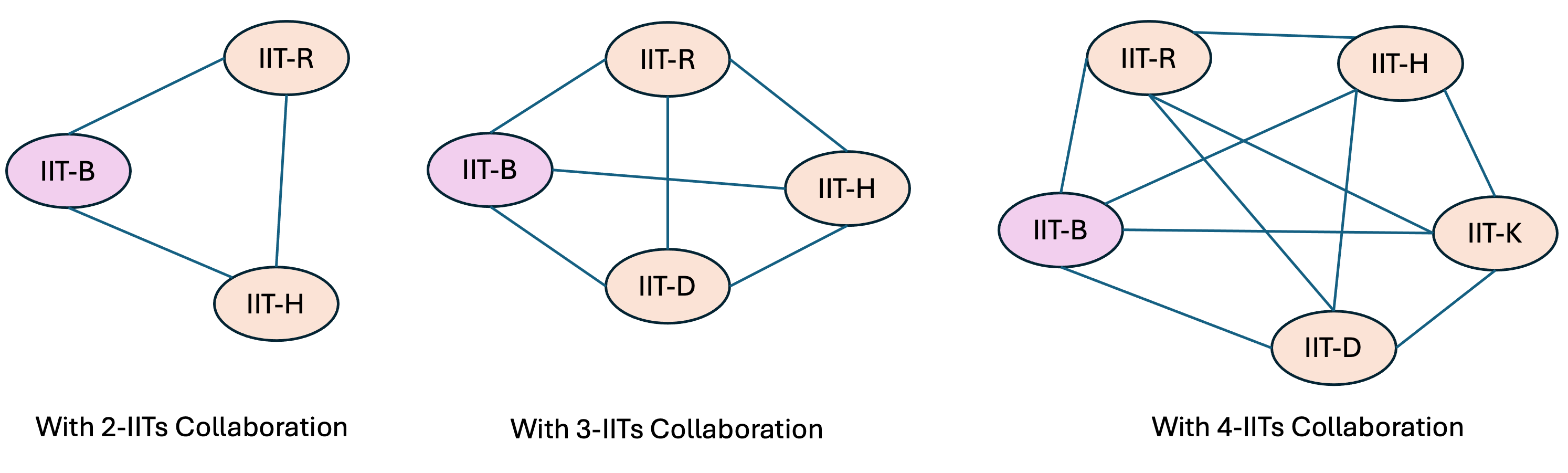} 
\caption{Demonstration of top IIT collaboration with other IITs.}
\label{fig:IITColabDemo}
\end{figure}

\begin{figure}[!h]
    \centering
\includegraphics[width=\linewidth]{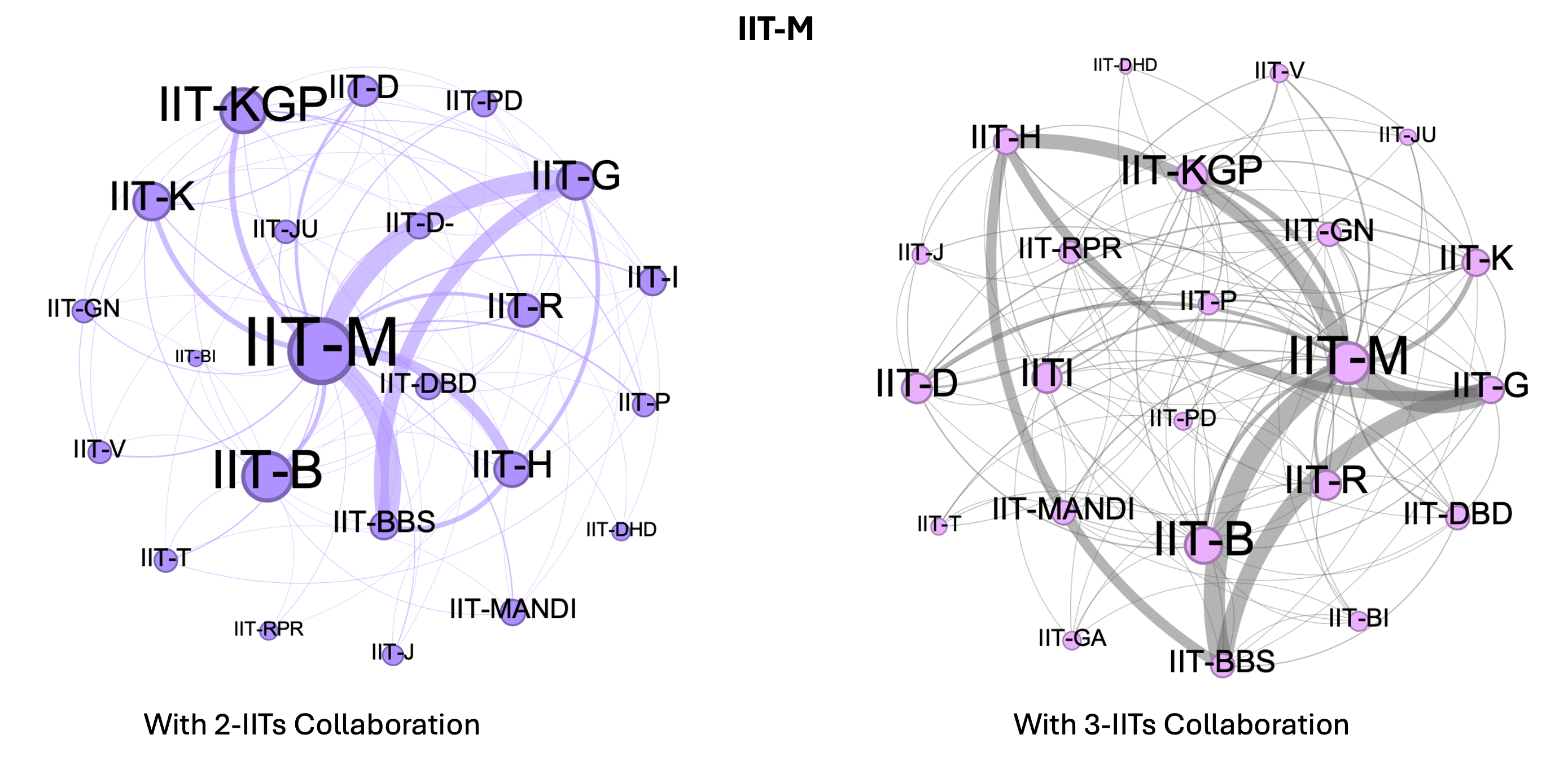} 
\caption{Collaboration network of IIT-M. (left) collaboration with two other IITs. The network has 23 nodes and 88 unique edges. (right) collaboration with three other IITs. The network has 23 nodes and 117 unique edges.}
\label{fig:IITMColab}
\end{figure}

Table~\ref{tab:IITColab} highlights the collaborative efforts among the leading IITs in India and shows trends in the research output at different levels of collaboration. The table displays the number of papers produced by each IIT in collaboration with other IITs, categorized by the number of collaborating institutes. Collaboration with 2 IITs represents papers produced when two other IITs collaborate with the main IIT in a single paper. Similarly, collaboration with 3 IITs represents papers produced when three other IITs collaborate with the main IIT in a single paper and so on (see Fig.~\ref{fig:IITColabDemo}). IIT-M consistently leads across all levels of collaboration, showing its strong emphasis on partnerships and its active role in collaborative research with other IITs. Most of the papers are the result of smaller collaborations involving two IITs, indicating that such partnerships are the most common and productive format for joint research, and the number of papers decreases as more IITs are involved in the collaboration. This trend suggests that smaller-scale partnerships are more practical and common in academic research among IITs. Figure~\ref{fig:IITMColab} shows the IIT-M collaboration network with 22 other IITs. The top collaborators of IIT-M are already shown in Figure~\ref{fig:InterIITColab}.

\begin{table}[!h]
\centering
\caption{Distribution of Research Papers Across Different Levels of Collaboration Among IITs.}
\begin{tabular}{|l|ccccc|}
\hline
\multirow{2}{*}{\textbf{Collaborations}} & \multicolumn{5}{c|}{\textbf{No. of Papers}}                \\ \cline{2-6} & \multicolumn{1}{c|}{\textbf{IIT-B}} & \multicolumn{1}{c|}{\textbf{IIT-D}} & \multicolumn{1}{c|}{\textbf{IIT-M}} & \multicolumn{1}{c|}{\textbf{IIT-KGP}} & \textbf{IIT-K} \\ \hline
\textbf{With 2 IITs}                     & \multicolumn{1}{c|}{153}            & \multicolumn{1}{c|}{118}            & \multicolumn{1}{c|}{224}            & \multicolumn{1}{c|}{179}              & 193            \\ \hline
\textbf{With 3 IITs}                     & \multicolumn{1}{c|}{22}             & \multicolumn{1}{c|}{21}             & \multicolumn{1}{c|}{85}             & \multicolumn{1}{c|}{31}               & 16             \\ \hline
\textbf{With 4 IITs}                     & \multicolumn{1}{c|}{8}              & \multicolumn{1}{c|}{6}              & \multicolumn{1}{c|}{7}              & \multicolumn{1}{c|}{5}                & 5              \\ \hline
\textbf{With 6 IITs}                     & \multicolumn{1}{c|}{1}              & \multicolumn{1}{c|}{0}              & \multicolumn{1}{c|}{1}              & \multicolumn{1}{c|}{0}                & 0              \\ \hline
\textbf{Total}                           & \multicolumn{1}{c|}{184}            & \multicolumn{1}{c|}{145}            & \multicolumn{1}{c|}{317}            & \multicolumn{1}{c|}{215}              & 214            \\ \hline
\end{tabular}
\label{tab:IITColab}
\end{table}

\subsection{Authors team size analysis}

Table~\ref{table:teamSize} shows the percentage distribution of authors by team size in five IITs. Teams of two authors make a significant contribution, with IIT-M (35. 18\%) and IIT-K (32. 57\%) being the highest contributors. IIT-D (29.98\%) and IIT-KGP (29.9\%) are close, while IIT-B has the smallest percentage for this category (28.5\%). The team comprising 3-5 authors is the most dominant category, with percentages exceeding 50\% for all IITs. IIT-D leads with 55.51\%, followed by IIT-M (50.85\%) and IIT-KGP (54.15\%). This suggests that most academic papers in these IITs are written collaboratively by mid-sized teams. Extremely large teams (100 authors) contribute a negligible percentage in all IITs, with values not exceeding 1\%. IIT-B (0.96\%) contributes the most in this category, followed by IIT-M (0.67\%).  In general, IIT-B appears to have the most diverse team size distribution.

\begin{table}[!h]
\centering
\caption{Number of papers (\%) as per number of authors per paper (team size).}
\begin{tabular}{|c|ccccc|}
\hline
\multirow{2}{*}{\textbf{Team Size}} & \multicolumn{5}{c|}{\textbf{Number of Authors (in \%)}}                                                                                                                  \\ \cline{2-6} 
                                    & \multicolumn{1}{c|}{\textbf{IIT-B}} & \multicolumn{1}{c|}{\textbf{IIT-D}} & \multicolumn{1}{c|}{\textbf{IIT-M}} & \multicolumn{1}{c|}{\textbf{IIT-KGP}} & \textbf{IIT-K} \\ \hline

2 Authors                          & \multicolumn{1}{c|}{28.5}           & \multicolumn{1}{c|}{29.98}          & \multicolumn{1}{c|}{35.18}          & \multicolumn{1}{c|}{29.9}             & 32.57          \\ \hline
3-5 Authors                         & \multicolumn{1}{c|}{52.52}          & \multicolumn{1}{c|}{55.51}          & \multicolumn{1}{c|}{50.85}          & \multicolumn{1}{c|}{54.15}            & 51.66          \\ \hline
6-10 Authors                        & \multicolumn{1}{c|}{11.75}          & \multicolumn{1}{c|}{9.05}           & \multicolumn{1}{c|}{8.57}           & \multicolumn{1}{c|}{10.24}            & 8.31           \\ \hline
11-20 Authors                       & \multicolumn{1}{c|}{1.24}           & \multicolumn{1}{c|}{1.01}           & \multicolumn{1}{c|}{0.83}           & \multicolumn{1}{c|}{0.9}              & 0.94           \\ \hline
100 Authors                         & \multicolumn{1}{c|}{0.96}           & \multicolumn{1}{c|}{0.02}           & \multicolumn{1}{c|}{0.67}           & \multicolumn{1}{c|}{0.02}             & 0.01           \\ \hline
\end{tabular}
\label{table:teamSize}
\end{table}
\subsection{Authorship position distribution}

\begin{figure}[!h]
    \centering
\includegraphics[width=0.75\linewidth]{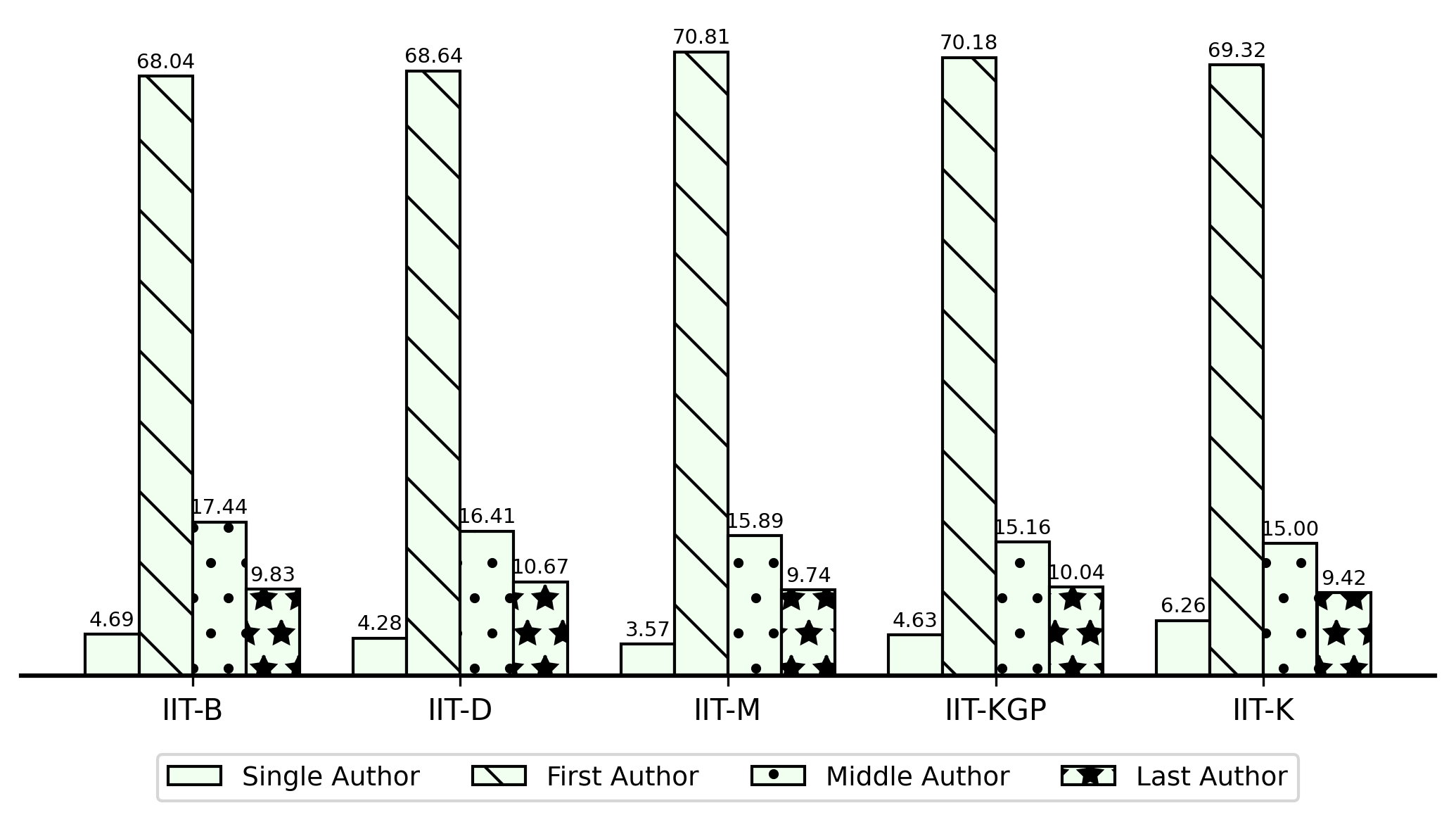} 
\caption{Distribution of papers based on authorship position.}
\label{fig:AuthorPosition}
\end{figure}

In research, it is often assumed that the first author plays a leading role in the research and writing process~\citep{persson2001all, zbar2011significance}.
Figure~\ref{fig:AuthorPosition} shows the distribution of contributions of authors in different authorship positions (e.g. single author, first author, middle author, last author) in five IITs. The largest contribution across all institutes is made by first authors and middle authors consistently contribute a moderate percentage (around 15\% to 17\% across institutes). The contribution of the Last Authors is relatively consistent across institutes, hovering around 9\% to 10\%. The single authored paper contributed by all IITs is a small proportion with values in the range of approximately 3\% to 6\%. Overall, the distribution of author positions highlights the hierarchical nature of academic contributions in these institutions, where the first authors play the most significant role.
\subsubsection{Top authors as per publications}
Table~\ref{table:topAuthors} shows the list of the top 10 authors across all five IITs. Prof. B.P. Singh from IIT-D has the highest number of publications among all top researchers followed by B. K. Panigrahi from IIT-D, also. Top two authors from IIT-B and IIT-D are from Electrical Engineering, IIT-M and IIT-K are from Chemistry and Electrical Engineering, respectively, and IIT-KGP from Mechanical Engineering and Materials Science \& Engineering.
\begin{table}[!h]
\centering
\caption{List of top 10 authors from five IITs based on number of publications.}
\begin{tabular}{|cllccclc|}
\hline
\multicolumn{8}{|c|}{\textbf{IIT-B}} \\ \hline
\multicolumn{1}{|c|}{\textbf{Rank}} & \multicolumn{1}{l|}{\textbf{Scopus ID}} & \multicolumn{1}{l|}{\textbf{Author Name}}                                                & \multicolumn{1}{c|}{\textbf{\#Papers}} & \multicolumn{1}{c|}{\textbf{Rank}} & \multicolumn{1}{c|}{\textbf{Author ID}} & \multicolumn{1}{l|}{\textbf{Author Name}}                                                  & \textbf{\#Papers} \\ \hline
\multicolumn{1}{|c|}{1}             & \multicolumn{1}{l|}{57192900310}         & \multicolumn{1}{l|}{Chakrabarti, S.}                                                     & \multicolumn{1}{c|}{405}               & \multicolumn{1}{c|}{6}             & \multicolumn{1}{c|}{26540761700}         & \multicolumn{1}{l|}{\begin{tabular}[c]{@{}l@{}}Kotha, \\ Sambasivarao\end{tabular}}        & 323               \\ \hline
\multicolumn{1}{|c|}{2}             & \multicolumn{1}{l|}{7402899220}         & \multicolumn{1}{l|}{Rao, V. Ramgopal}                                                    & \multicolumn{1}{c|}{366}               & \multicolumn{1}{c|}{7}             & \multicolumn{1}{c|}{35462660300}         & \multicolumn{1}{l|}{Singh, T. N.}                                                          & 310               \\ \hline
\multicolumn{1}{|c|}{3}             & \multicolumn{1}{l|}{7101803108}         & \multicolumn{1}{l|}{Bhattacharyya, P.}                                                   & \multicolumn{1}{c|}{347}               & \multicolumn{1}{c|}{8}             & \multicolumn{1}{c|}{9635357300}         & \multicolumn{1}{l|}{Baghini, M. S.}                                                        & 309               \\ \hline
\multicolumn{1}{|c|}{4}             & \multicolumn{1}{l|}{57204083493}         & \multicolumn{1}{l|}{Agarwal, Vivek}                                                      & \multicolumn{1}{c|}{344}               & \multicolumn{1}{c|}{9}             & \multicolumn{1}{c|}{7003781658}         & \multicolumn{1}{l|}{Lahiri, G. K.}                                                         & 301               \\ \hline
\multicolumn{1}{|c|}{5}             & \multicolumn{1}{l|}{7003788650}         & \multicolumn{1}{l|}{Ravikanth, M.}                                                       & \multicolumn{1}{c|}{336}               & \multicolumn{1}{c|}{10}            & \multicolumn{1}{c|}{7006039039}         & \multicolumn{1}{l|}{Bahadur, D.}                                                           & 292               \\ \hline
\multicolumn{8}{|c|}{\textbf{IIT-D}}  \\ \hline
\multicolumn{1}{|c|}{\textbf{Rank}} & \multicolumn{1}{l|}{\textbf{Author ID}} & \multicolumn{1}{l|}{\textbf{Author Name}}                                                & \multicolumn{1}{c|}{\textbf{\#Papers}} & \multicolumn{1}{c|}{\textbf{Rank}} & \multicolumn{1}{c|}{\textbf{Author ID}} & \multicolumn{1}{l|}{\textbf{Author Name}}                                                  & \textbf{\#Papers} \\ \hline
\multicolumn{1}{|c|}{1}             & \multicolumn{1}{l|}{7405638726}         & \multicolumn{1}{l|}{Singh, B. P.}                                                        & \multicolumn{1}{c|}{2363}              & \multicolumn{1}{c|}{6}             & \multicolumn{1}{c|}{7005406958}         & \multicolumn{1}{l|}{Rathore, A. S.}                                                        & 360               \\ \hline
\multicolumn{1}{|c|}{2}             & \multicolumn{1}{l|}{25637555100}         & \multicolumn{1}{l|}{Panigrahi, B. K.}                                                    & \multicolumn{1}{c|}{746}               & \multicolumn{1}{c|}{7}             & \multicolumn{1}{c|}{36067733400}         & \multicolumn{1}{l|}{Sharma, R. P.}                                                         & 353               \\ \hline
\multicolumn{1}{|c|}{3}             & \multicolumn{1}{l|}{56606361700}         & \multicolumn{1}{l|}{Mishra, S.}                                                          & \multicolumn{1}{c|}{428}               & \multicolumn{1}{c|}{8}             & \multicolumn{1}{c|}{7103015030}         & \multicolumn{1}{l|}{Tripathi, V. K.}                                                       & 335               \\ \hline
\multicolumn{1}{|c|}{4}             & \multicolumn{1}{l|}{55849312100}         & \multicolumn{1}{l|}{Koul, Shiban K.}                                                     & \multicolumn{1}{c|}{423}               & \multicolumn{1}{c|}{9}             & \multicolumn{1}{c|}{7006203679}         & \multicolumn{1}{l|}{Ganguli, Ashok K.}                                                     & 322               \\ \hline
\multicolumn{1}{|c|}{5}             & \multicolumn{1}{l|}{7202949674}         & \multicolumn{1}{l|}{Kaushik, S. C.}                                                      & \multicolumn{1}{c|}{371}               & \multicolumn{1}{c|}{10}            & \multicolumn{1}{c|}{55450482500}          & \multicolumn{1}{l|}{Tiwari, G. N.}                                                         & 315               \\ \hline
\multicolumn{8}{|c|}{\textbf{IIT-M}} \\ \hline
\multicolumn{1}{|c|}{\textbf{Rank}} & \multicolumn{1}{l|}{\textbf{Author ID}} & \multicolumn{1}{l|}{\textbf{Author Name}}                                                & \multicolumn{1}{c|}{\textbf{\#Papers}} & \multicolumn{1}{c|}{\textbf{Rank}} & \multicolumn{1}{c|}{\textbf{Author ID}} & \multicolumn{1}{l|}{\textbf{Author Name}}                                                  & \textbf{\#Papers} \\ \hline
\multicolumn{1}{|c|}{1}             & \multicolumn{1}{l|}{7006641708}         & \multicolumn{1}{l|}{Pradeep, T.}                                                         & \multicolumn{1}{c|}{509}               & \multicolumn{1}{c|}{6}             & \multicolumn{1}{c|}{9535936500}         & \multicolumn{1}{l|}{Ramaprabhu, S.}                                                        & 350               \\ \hline
\multicolumn{1}{|c|}{2}             & \multicolumn{1}{l|}{7003779585}         & \multicolumn{1}{l|}{Sarathi, R.}                                                         & \multicolumn{1}{c|}{383}               & \multicolumn{1}{c|}{7}             & \multicolumn{1}{c|}{7102087090}         & \multicolumn{1}{l|}{\begin{tabular}[c]{@{}l@{}}Murthy, C. \\ Siva Ram\end{tabular}}        & 348               \\ \hline
\multicolumn{1}{|c|}{3}             & \multicolumn{1}{l|}{7102925688}         & \multicolumn{1}{l|}{Viswanathan, B.}                                                     & \multicolumn{1}{c|}{373}               & \multicolumn{1}{c|}{8}             & \multicolumn{1}{c|}{6701351287}         & \multicolumn{1}{l|}{\begin{tabular}[c]{@{}l@{}}Rao, M. S.\\ Ramachandra\end{tabular}}      & 319               \\ \hline
\multicolumn{1}{|c|}{4}             & \multicolumn{1}{l|}{35551753500}         & \multicolumn{1}{l|}{Murty, B. S.}                                                        & \multicolumn{1}{c|}{365}               & \multicolumn{1}{c|}{9}             & \multicolumn{1}{c|}{20734150400}         & \multicolumn{1}{l|}{Mukesh, D.}                                                            & 294               \\ \hline
\multicolumn{1}{|c|}{5}             & \multicolumn{1}{l|}{7005602178}         & \multicolumn{1}{l|}{\begin{tabular}[c]{@{}l@{}}Balasubramaniam, \\ K. K.\end{tabular}}   & \multicolumn{1}{c|}{359}               & \multicolumn{1}{c|}{10}            & \multicolumn{1}{c|}{6506470218}         & \multicolumn{1}{l|}{\begin{tabular}[c]{@{}l@{}}Sivaprakasam, \\ Mohanasankar\end{tabular}} & 291               \\ \hline
\multicolumn{8}{|c|}{\textbf{IIT-KGP}}  \\ \hline
\multicolumn{1}{|c|}{\textbf{Rank}} & \multicolumn{1}{l|}{\textbf{Author ID}} & \multicolumn{1}{l|}{\textbf{Author Name}}                                                & \multicolumn{1}{c|}{\textbf{\#Papers}} & \multicolumn{1}{c|}{\textbf{Rank}} & \multicolumn{1}{c|}{\textbf{Author ID}} & \multicolumn{1}{l|}{\textbf{Author Name}}                                                  & \textbf{\#Papers} \\ \hline
\multicolumn{1}{|c|}{1}             & \multicolumn{1}{l|}{56643797300}         & \multicolumn{1}{l|}{Chakraborty, S.}                                                     & \multicolumn{1}{c|}{550}               & \multicolumn{1}{c|}{6}             & \multicolumn{1}{c|}{56269528000}          & \multicolumn{1}{l|}{Chattaraj, P. K.}                                                      & 392               \\ \hline
\multicolumn{1}{|c|}{2}             & \multicolumn{1}{l|}{26643264700}         & \multicolumn{1}{l|}{Bhowmick, Anil K.}                                                   & \multicolumn{1}{c|}{540}               & \multicolumn{1}{c|}{7}             & \multicolumn{1}{c|}{7202304528}         & \multicolumn{1}{l|}{De, Sirshendu}                                                         & 354               \\ \hline
\multicolumn{1}{|c|}{3}             & \multicolumn{1}{l|}{7401768547}         & \multicolumn{1}{l|}{Misra, Sudip}                                                        & \multicolumn{1}{c|}{485}               & \multicolumn{1}{c|}{8}             & \multicolumn{1}{c|}{58722066100}         & \multicolumn{1}{l|}{Mukherjee, Jayanta}                                                    & 299               \\ \hline
\multicolumn{1}{|c|}{4}             & \multicolumn{1}{l|}{57203479999}         & \multicolumn{1}{l|}{Ray, S. K.}                                                          & \multicolumn{1}{c|}{484}               & \multicolumn{1}{c|}{9}             & \multicolumn{1}{c|}{24824941700}         & \multicolumn{1}{l|}{Pal, T.}                                                               & 299               \\ \hline
\multicolumn{1}{|c|}{5}             & \multicolumn{1}{l|}{7102410371}         & \multicolumn{1}{l|}{Choudhary, R. N.P.}                                                  & \multicolumn{1}{c|}{408}               & \multicolumn{1}{c|}{10}            & \multicolumn{1}{c|}{55927861800}         & \multicolumn{1}{l|}{Routray, Aurobinda}                                                    & 297               \\ \hline
\multicolumn{8}{|c|}{\textbf{IIT-K}}  \\ \hline
\multicolumn{1}{|c|}{\textbf{Rank}} & \multicolumn{1}{l|}{\textbf{Author ID}} & \multicolumn{1}{l|}{\textbf{Author Name}}                                                & \multicolumn{1}{c|}{\textbf{\#Papers}} & \multicolumn{1}{c|}{\textbf{Rank}} & \multicolumn{1}{c|}{\textbf{Author ID}} & \multicolumn{1}{l|}{\textbf{Author Name}}                                                  & \textbf{\#Papers} \\ \hline
\multicolumn{1}{|c|}{1}             & \multicolumn{1}{l|}{7006003609}         & \multicolumn{1}{l|}{\begin{tabular}[c]{@{}l@{}}Chandrasekhar, \\ Vadapalli\end{tabular}} & \multicolumn{1}{c|}{362}               & \multicolumn{1}{c|}{6}             & \multicolumn{1}{c|}{8711683200}         & \multicolumn{1}{l|}{\begin{tabular}[c]{@{}l@{}}Srivastava, \\ Kumar Vaibhav\end{tabular}}  & 308               \\ \hline
\multicolumn{1}{|c|}{2}             & \multicolumn{1}{l|}{57201754568}         & \multicolumn{1}{l|}{Sharma, Ashutosh}                                                    & \multicolumn{1}{c|}{358}               & \multicolumn{1}{c|}{7}             & \multicolumn{1}{c|}{7202577198}         & \multicolumn{1}{l|}{Akhtar, M. J.}                                                         & 307               \\ \hline
\multicolumn{1}{|c|}{3}             & \multicolumn{1}{l|}{35479531700}          & \multicolumn{1}{l|}{\begin{tabular}[c]{@{}l@{}}Agarwal, \\ Avinash Kumar\end{tabular}}   & \multicolumn{1}{c|}{350}               & \multicolumn{1}{c|}{8}             & \multicolumn{1}{c|}{55740373500}          & \multicolumn{1}{l|}{Singh, SN}                                                             & 295               \\ \hline
\multicolumn{1}{|c|}{4}             & \multicolumn{1}{l|}{14029622100}          & \multicolumn{1}{l|}{Chauhan, Yogesh Singh}                                               & \multicolumn{1}{c|}{327}               & \multicolumn{1}{c|}{9}             & \multicolumn{1}{c|}{56657757000}         & \multicolumn{1}{l|}{Chhabra, R. P.}                                                        & 293               \\ \hline
\multicolumn{1}{|c|}{5}             & \multicolumn{1}{l|}{7005150925}         & \multicolumn{1}{l|}{Kundu, Debasis}                                                      & \multicolumn{1}{c|}{318}               & \multicolumn{1}{c|}{10}            & \multicolumn{1}{c|}{7006019904}         & \multicolumn{1}{l|}{Deb, Kalyanmoy}                                                        & 271               \\ \hline
\end{tabular}
\label{table:topAuthors}
\end{table}

\subsection{Country collaboration}

\begin{figure}[!h]
    \centering
\includegraphics[width=\linewidth]{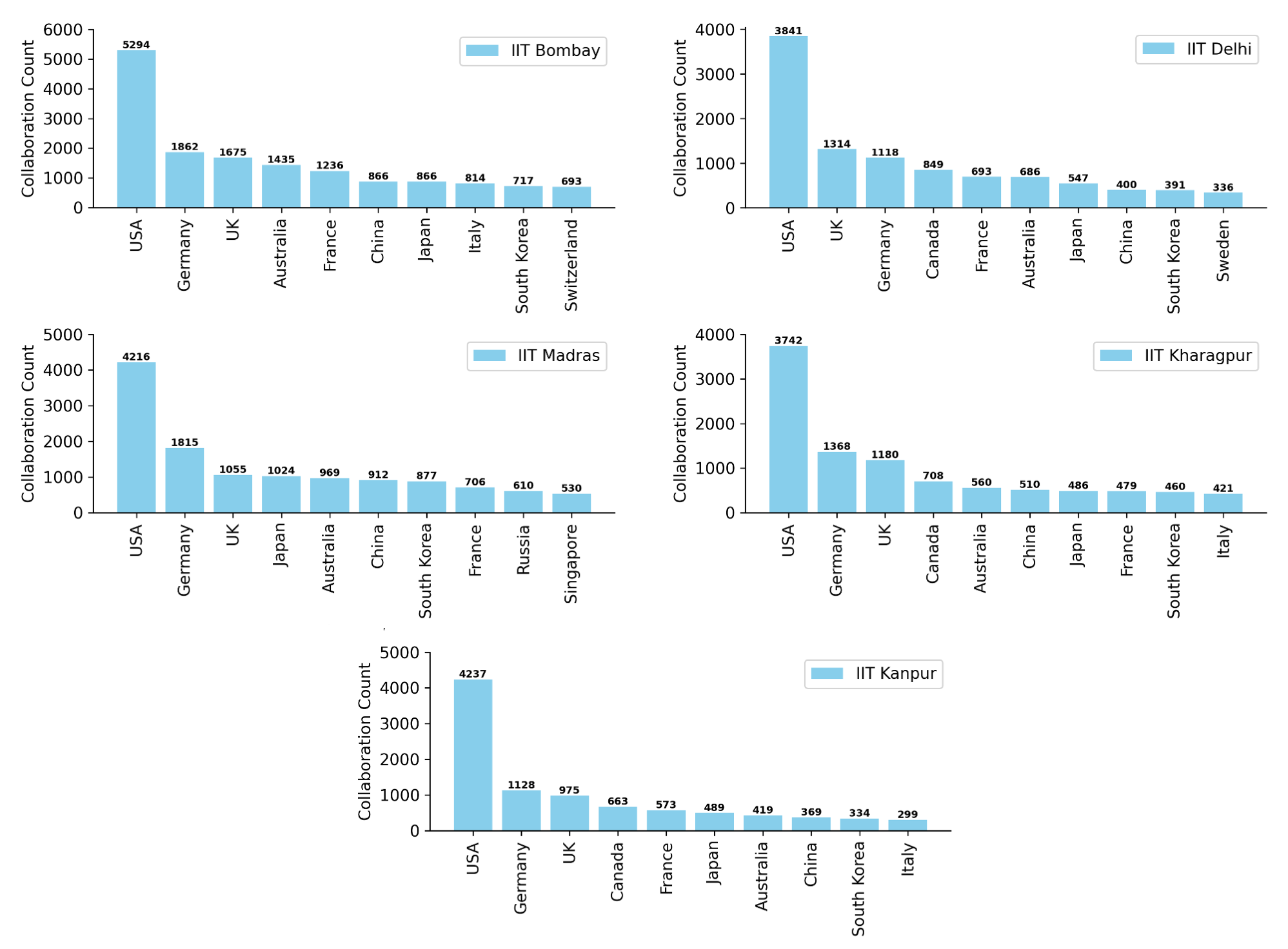} 
\caption{Country collaboration of five ITTs.}
\label{fig:countryColab}
\end{figure}

Figure~\ref{fig:countryColab} shows the bar charts of international collaboration counts between five IITs and other top collaborating countries. Each bar graph highlights the number of collaborative research publications between each IIT and different countries, sorted by the most frequent collaborators. The USA is a major collaborator, significantly more than other countries, which suggests that IIT-Bombay has a strong relationship with American institutions. Europe also has a strong presence, especially Germany and the UK. IIT-Delhi also has strong collaborations with the USA, UK, and Germany, although the volume with the USA is less than that of IIT-Bombay. The collaborations are similarly diverse, covering a wide range of countries in Europe, North America, and Asia. IIT-Madras has a strong link with the USA, and Germany, but there is also considerable collaboration with Japan and China, which may indicate research areas where collaboration with Asian countries is important. Unlike other IITs, IIT-Kharagpur has a strong preference for collaboration with the USA. The distribution is slightly less diverse, with fewer collaborations in East Asian countries compared to other IITs like IIT-Madras. IIT-Kanpur shows a similar pattern, with the USA as the primary collaborator, followed by Germany and the UK. East Asian countries have lower collaboration counts, which may indicate a focus on collaborations with North American and European countries.

Overall, the USA is the top collaborator for all IITs, with counts significantly higher than those for other countries, showing a strong reliance on American research institutions and funding sources for collaborative projects. Germany and the UK consistently appear among the top three collaborators for each IIT, suggesting a strong relationship with European institutions. Australia, Japan, and China are important collaborators, particularly for IIT-Bombay, IIT-Madras, and IIT-Delhi, indicating their participation in diverse international research networks. Collaborations with countries such as South Korea, Italy, and Russia vary between IITs, with some showing relatively low counts, especially IIT-Kanpur and IIT-Kharagpur. Each IIT shows some variation in its international collaboration patterns, but the overall trend emphasizes North American and European partnerships, with emerging collaborations in Asia. This pattern may reflect differences in research funding, faculty exchanges, and focus areas across IITs.

\subsubsection{Proportion of authorship position in top 3 countries}

\begin{figure}[!h]
    \centering
\includegraphics[width=\linewidth]{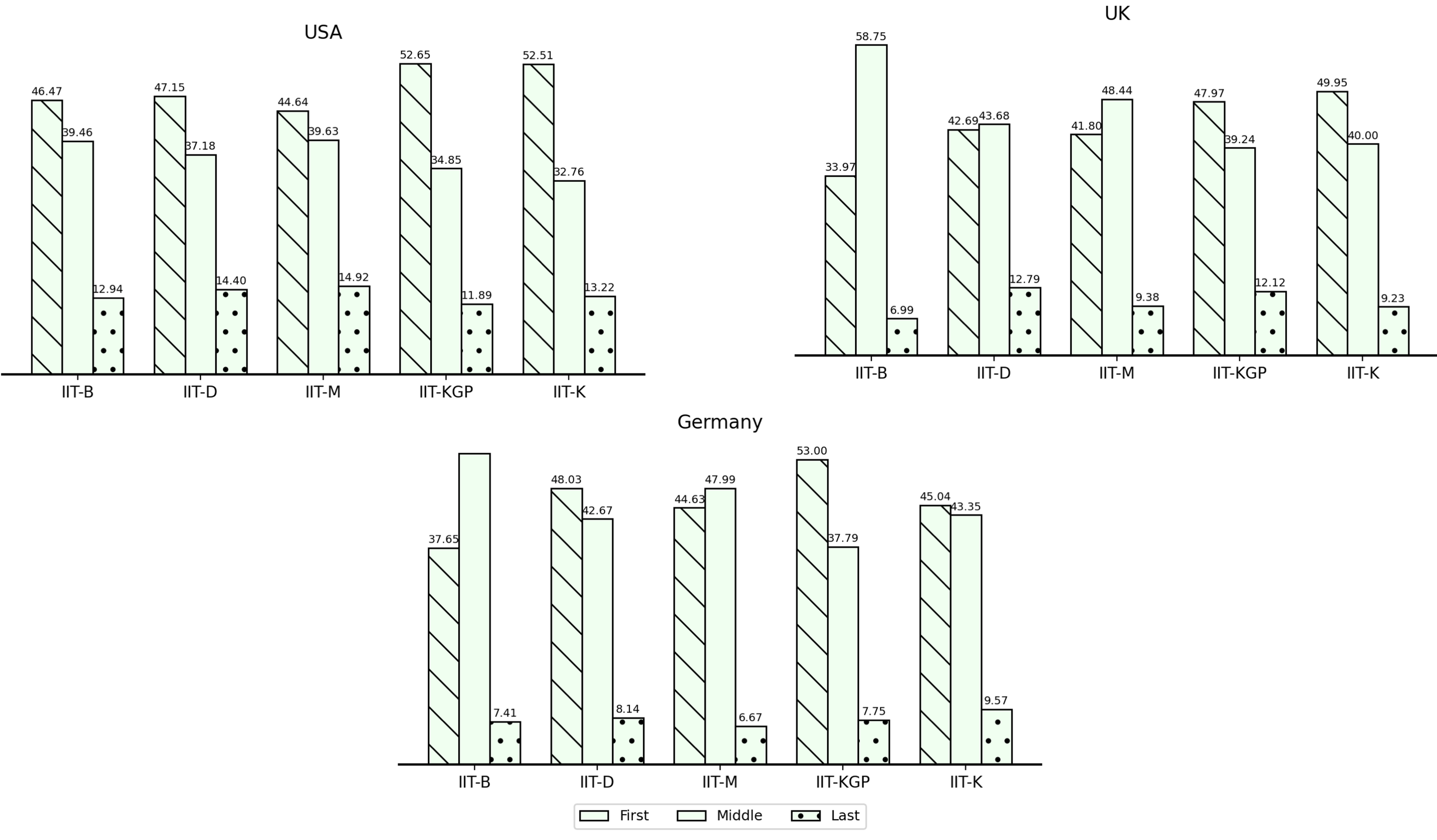} 
\caption{Distribution of authorship position for top three country collaborators.}
\label{fig:SDG_Country}
\end{figure}


\subsubsection{Collaboration in different regions and continents}

\begin{figure}[!h]
    \centering
\includegraphics[width=\linewidth]{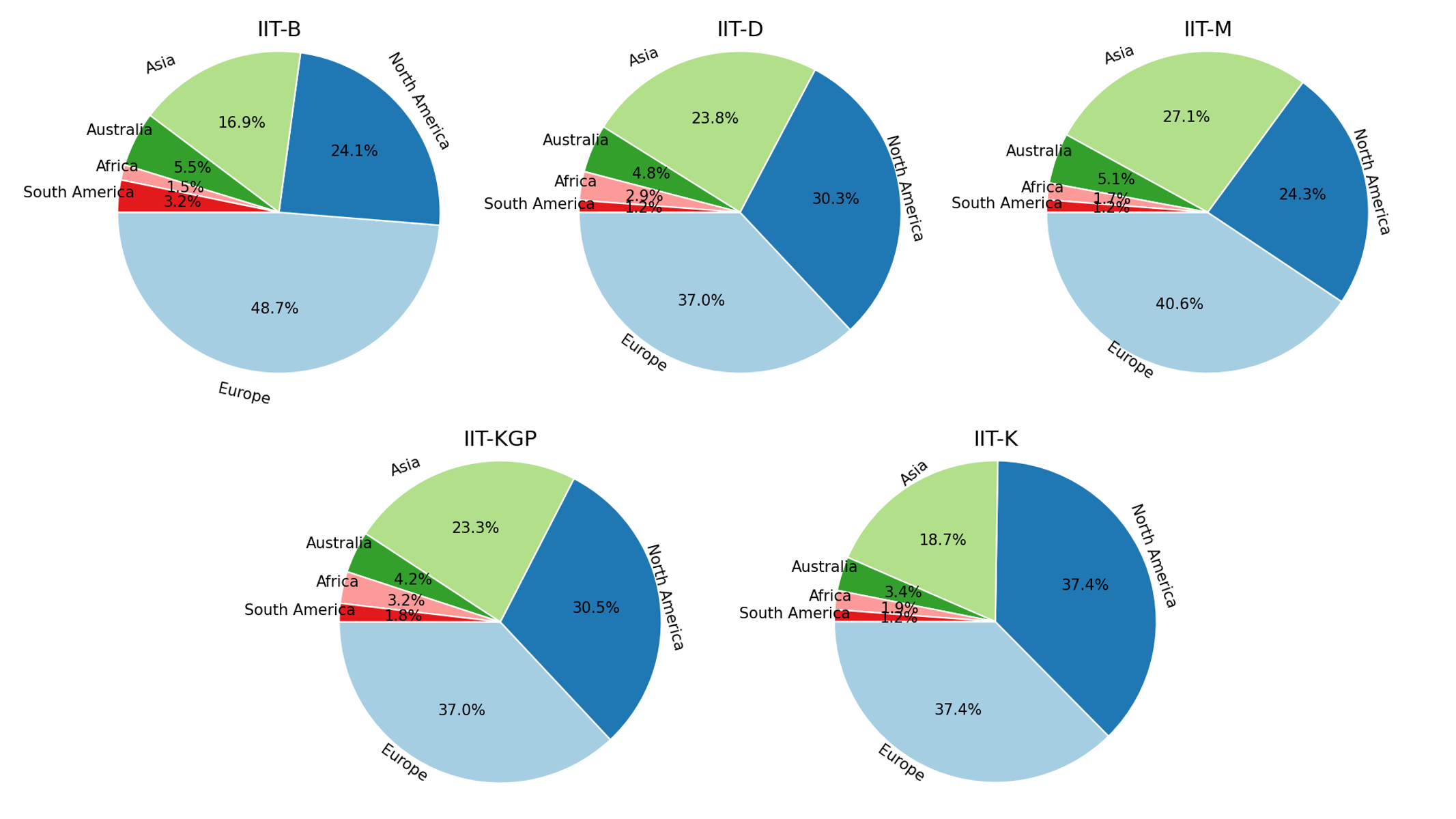} 
\caption{Research output distribution by continent across 5 IITs.}
\label{fig:Continent}
\end{figure}

Figure~\ref{fig:Continent} shows the distribution of research output across continents that shows consistent patterns among the five IITs, with Europe dominating collaborations. IIT-B shows the highest collaboration with Europe at 48.7\%, followed by IIT-M (40. 6\%), IIT-K (37. 4\%), IIT-D (37\%) and IIT-KGP (37\%). North America plays a secondary role in collaborations, contributing around 24–30\% for most IITs, with IIT-K and IIT-D showing slightly higher engagement. Asia emerges as the third significant collaborator, with IIT-M (27.1\%) and IIT-D (23.8\%) exhibiting the highest partnerships in this region. Australia, Africa, and South America collectively account for smaller percentages, typically less than 5\% each, reflecting existing but limited partnerships in these regions. IIT-B and IIT-M show slightly stronger collaborations with South America and Australia compared to the other IITs. This comparative analysis highlights Europe and North America as dominant contributors to IITs' research networks while pointing to potential growth opportunities in Asia and emerging collaborations in underrepresented regions like Africa and South America.

\subsection{Funding analysis}

Funding is essential for academic and scientific growth, especially in research-intensive institutions such as IITs. Funding drives research growth by enabling advanced infrastructure, attracting talent, supporting innovation, fostering collaborations, and improving research quality. It benefits society by addressing critical issues and boosts institutional reputation, creating a cycle of improvement and impact.
Table~\ref{table:funding} shows results on the research output supported by various prominent funding bodies across five IITs). The table provides the number of papers funded by each organization at each IIT and the total number of citations received by the papers funded by each organization at each IIT. Each IIT has different levels of support from various funding agencies, reflecting their specific research focuses. DST India and IIT internal funding are the most prominent funding sources, with the highest research output and impact across all IITs. IIT-KGP shows strong collaboration with NSF, leading to high impact, while other IITs vary in NSF-supported research. Funding bodies like CSIR, SERB, and UGC provide broad and consistent support across IITs, while bodies like MEITY and DRDO have smaller impacts.

\begin{table}[!h]
\centering
\caption{List of prominent funding bodies in India.}
\begin{tabular}{|l|cc|cc|cc|cc|cc|}
\hline
\multirow{2}{*}{\textbf{\begin{tabular}[c]{@{}l@{}}Funding \\ Bodies\end{tabular}}} & \multicolumn{2}{c|}{\textbf{IIT-B}}                  & \multicolumn{2}{c|}{\textbf{IIT-D}}                  & \multicolumn{2}{c|}{\textbf{IIT-M}}                  & \multicolumn{2}{c|}{\textbf{IIT-KGP}}                & \multicolumn{2}{c|}{\textbf{IIT-K}}                  \\ \cline{2-11} 
& \multicolumn{1}{c|}{\textbf{\#Papers}} & \textbf{TC} & \multicolumn{1}{c|}{\textbf{\#Papers}} & \textbf{TC} & \multicolumn{1}{c|}{\textbf{\#Papers}} & \textbf{TC} & \multicolumn{1}{c|}{\textbf{\#Papers}} & \textbf{TC} & \multicolumn{1}{c|}{\textbf{\#Papers}} & \textbf{TC} \\ \hline
\textbf{DST India}                                                                  & \multicolumn{1}{c|}{2851}              & 43262       & \multicolumn{1}{c|}{3084}              & 49121       & \multicolumn{1}{c|}{2885}              & 47621       & \multicolumn{1}{c|}{2688}              & 50037       & \multicolumn{1}{c|}{2163}              & 45432       \\ \hline
\textbf{IIT}                                                                        & \multicolumn{1}{c|}{1223}              & 15077       & \multicolumn{1}{c|}{751}               & 12140       & \multicolumn{1}{c|}{1400}              & 15492       & \multicolumn{1}{c|}{1244}              & 21622       & \multicolumn{1}{c|}{1663}              & 22909       \\ \hline
\textbf{CSIR}                                                                       & \multicolumn{1}{c|}{1080}              & 20475       & \multicolumn{1}{c|}{1259}              & 32985       & \multicolumn{1}{c|}{721}               & 12909       & \multicolumn{1}{c|}{1182}              & 31431       & \multicolumn{1}{c|}{682}               & 14472       \\ \hline
\textbf{NSF}                                                                        & \multicolumn{1}{c|}{745}               & 45618       & \multicolumn{1}{c|}{394}               & 15319       & \multicolumn{1}{c|}{602}               & 25161       & \multicolumn{1}{c|}{296}               & 9689        & \multicolumn{1}{c|}{370}               & 9792        \\ \hline
\textbf{SERB}                                                                       & \multicolumn{1}{c|}{735}               & 7196        & \multicolumn{1}{c|}{761}               & 8971        & \multicolumn{1}{c|}{606}               & 5867        & \multicolumn{1}{c|}{761}               & 10233       & \multicolumn{1}{c|}{721}               & 7366        \\ \hline
\textbf{DBT}                                                                        & \multicolumn{1}{c|}{383}               & 6127        & \multicolumn{1}{c|}{538}               & 11993       & \multicolumn{1}{c|}{294}               & 4270        & \multicolumn{1}{c|}{342}               & 14573       & \multicolumn{1}{c|}{151}               & 3006        \\ \hline
\textbf{DST Kerala}                                                                 & \multicolumn{1}{c|}{340}               & 7595        & \multicolumn{1}{c|}{388}               & 10366       & \multicolumn{1}{c|}{357}               & 9037        & \multicolumn{1}{c|}{306}               & 10573       & \multicolumn{1}{c|}{368}               & 11258       \\ \hline
\textbf{UGC}                                                                        & \multicolumn{1}{c|}{324}               & 3598        & \multicolumn{1}{c|}{539}               & 9099        & \multicolumn{1}{c|}{258}               & 3383        & \multicolumn{1}{c|}{144}               & 4128        & \multicolumn{1}{c|}{152}               & 2183        \\ \hline
\textbf{MoE}                                                                        & \multicolumn{1}{c|}{222}               & 1575        & \multicolumn{1}{c|}{264}               & 1540        & \multicolumn{1}{c|}{247}               & 1105        & \multicolumn{1}{c|}{431}               & 2930        & \multicolumn{1}{c|}{91}                & 498         \\ \hline
\textbf{MHRD}                                                                       & \multicolumn{1}{c|}{219}               & 4575        & \multicolumn{1}{c|}{225}               & 5827        & \multicolumn{1}{c|}{204}               & 3937        & \multicolumn{1}{c|}{567}               & 14682       & \multicolumn{1}{c|}{141}               & 42042       \\ \hline
\textbf{MNRE}                                                                       & \multicolumn{1}{c|}{204}               & 4250        & \multicolumn{1}{c|}{69}                & 2668        & \multicolumn{1}{c|}{35}                & 1489        & \multicolumn{1}{c|}{52}                & 1796        & \multicolumn{1}{c|}{41}                & 1058        \\ \hline
\textbf{ISRO}                                                                       & \multicolumn{1}{c|}{172}               & 1893        & \multicolumn{1}{c|}{31}                & 221         & \multicolumn{1}{c|}{68}                & 867         & \multicolumn{1}{c|}{132}               & 2486        & \multicolumn{1}{c|}{120}               & 2163        \\ \hline
\textbf{DAE}                                                                        & \multicolumn{1}{c|}{166}               & 2096        & \multicolumn{1}{c|}{73}                & 1032        & \multicolumn{1}{c|}{102}               & 1913        & \multicolumn{1}{c|}{142}               & 3348        & \multicolumn{1}{c|}{125}               & 1988        \\ \hline
\textbf{MEITy}                                                                      & \multicolumn{1}{c|}{144}               & 672         & \multicolumn{1}{c|}{97}                & 1048        & \multicolumn{1}{c|}{86}                & 422         & \multicolumn{1}{c|}{94}                & 426         & \multicolumn{1}{c|}{104}               & 545         \\ \hline
\textbf{DRDO}                                                                       & \multicolumn{1}{c|}{107}               & 1371        & \multicolumn{1}{c|}{216}               & 3062        & \multicolumn{1}{c|}{235}               & 4915        & \multicolumn{1}{c|}{187}               & 3704        & \multicolumn{1}{c|}{100}               & 2571        \\ \hline
\end{tabular}
\label{table:funding}
\end{table}


\subsection{SDGs contribution}

Research on SDGs is essential because it addresses global challenges such as poverty, inequality, and climate change by providing evidence-based solutions and driving innovation for sustainability. It informs policies, fosters interdisciplinary collaboration, and helps monitor progress toward these goals. By guiding resource allocation and creating resilience and equality strategies, SDG research ensures a sustainable future for all~\citep{sachs2019six,assembly2015resolution}.
Table~\ref{tab:SDG_count} provides data on the research output of five Indian Institutes of Technology (IIT), specifically analyzing the extent to which their articles are mapped to the Sustainable Development Goals (SDGs). IIT-D and IIT-KGP perform better in aligning their research output with SDGs compared to other IITs. IIT-K lags behind in SDG mapping, with the lowest percentage of mapped papers.
Across all IITs, the average percentage of papers mapped to SDGs is approximately 35\%, indicating a significant gap in research alignment with SDGs. While the IITs are contributing significantly to research, there is substantial room for improvement in mapping their output to SDGs. Encouraging more research initiatives focused on sustainability could enhance alignment with global development goals.

\begin{table}[!h]
\centering
\caption{SDG count as per each IIT. The values colored in green highlighting the major contributions. Darker the color higher the contribution and vice versa.}
\begin{tabular}{|l|c|cccc|}
\hline
\multirow{3}{*}{\textbf{IIT}} & \multirow{3}{*}{\textbf{Total Papers}} & \multicolumn{4}{c|}{\textbf{Mapped   with SDG}}                                                                                  \\ \cline{3-6}  &                                        & \multicolumn{2}{c|}{\textbf{Yes}}                                         & \multicolumn{2}{c|}{\textbf{No}}                     \\ \cline{3-6} 
  &  & \multicolumn{1}{c|}{\textbf{Count}} & \multicolumn{1}{c|}{\textbf{in \%}} & \multicolumn{1}{c|}{\textbf{Count}} & \textbf{in \%} \\ \hline
IIT-B                         & 47783                                  & \multicolumn{1}{c|}{16781}          & \multicolumn{1}{c|}{35.12}          & \multicolumn{1}{c|}{31002}          & 64.88          \\ \hline
IIT-D                         & 53547                                  & \multicolumn{1}{c|}{21305}          & \multicolumn{1}{c|}{39.79}          & \multicolumn{1}{c|}{32242}          & 60.21          \\ \hline
IIT-M                         & 50025                                  & \multicolumn{1}{c|}{16138}          & \multicolumn{1}{c|}{32.26}          & \multicolumn{1}{c|}{33887}          & 67.74          \\ \hline
IIT-KGP                       & 53232                                  & \multicolumn{1}{c|}{19834}          & \multicolumn{1}{c|}{37.26}          & \multicolumn{1}{c|}{33398}          & 62.74          \\ \hline
IIT-K                         & 38584                                  & \multicolumn{1}{c|}{10984}          & \multicolumn{1}{c|}{28.47}          & \multicolumn{1}{c|}{27600}          & 71.53          \\ \hline
Total                         & 243171                                 & \multicolumn{1}{c|}{85042}          & \multicolumn{1}{c|}{34.97}          & \multicolumn{1}{c|}{158129}         & 65.03          \\ \hline
\end{tabular}
\label{tab:SDG_count}
\end{table}

\begin{figure}[!h]
    \centering
\includegraphics[width=0.85\linewidth]{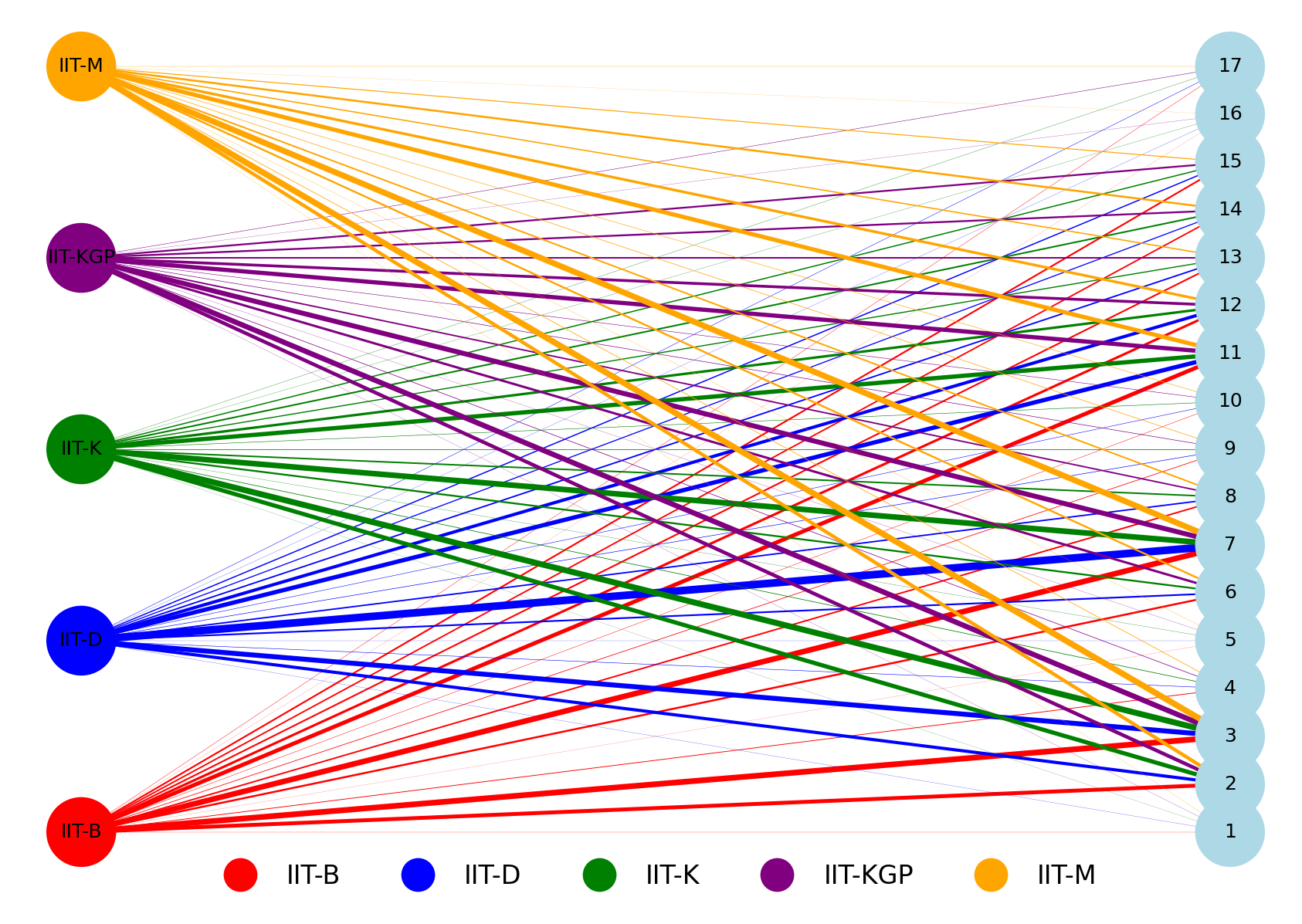} 
\caption{Bipartite graph of SDG orientation of five IITs.}
\label{fig:SDGgraph}
\end{figure}
\begin{figure}[!h]
    \centering
\includegraphics[width=0.85\linewidth]{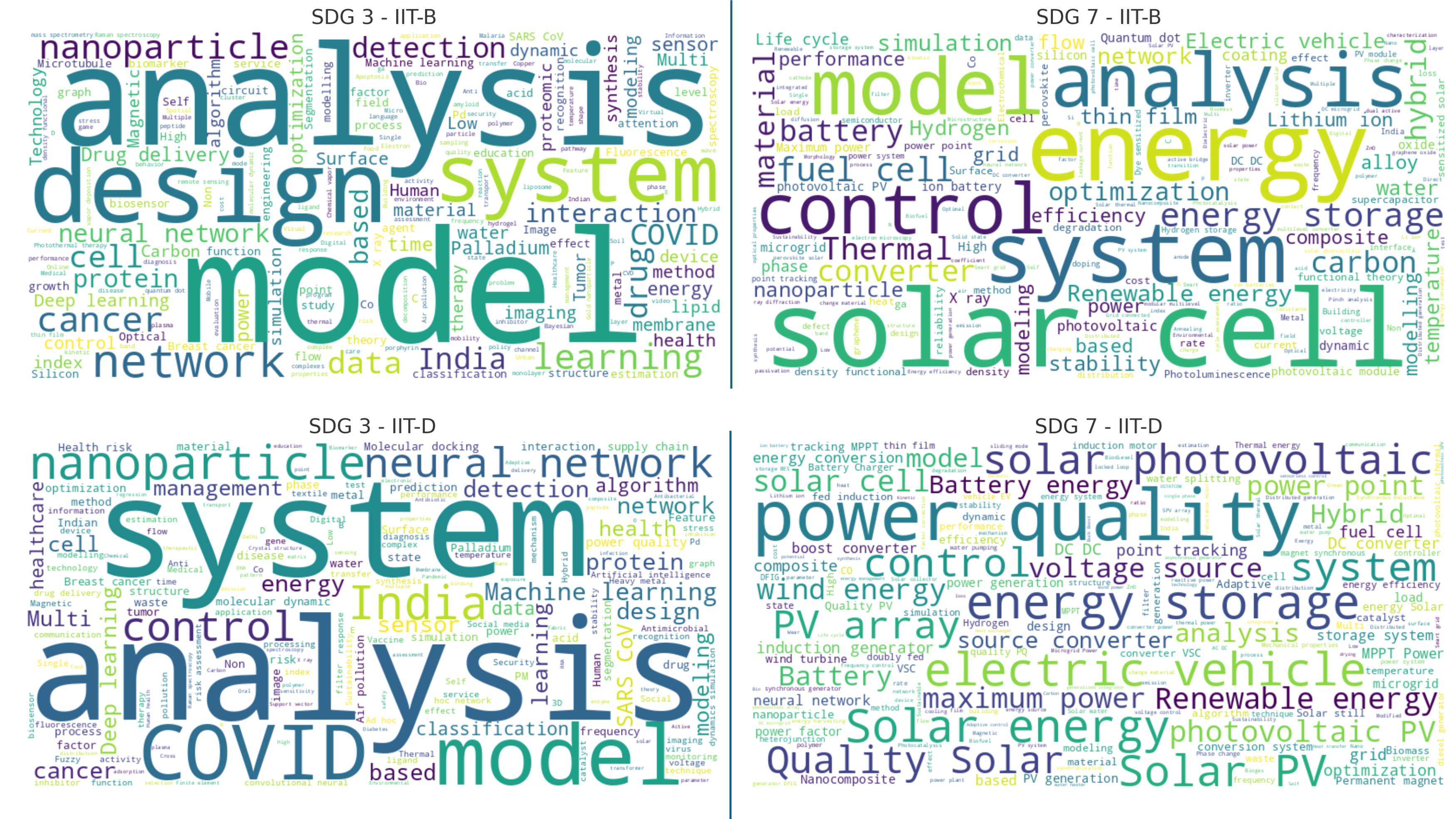} 
\includegraphics[width=0.85\linewidth]{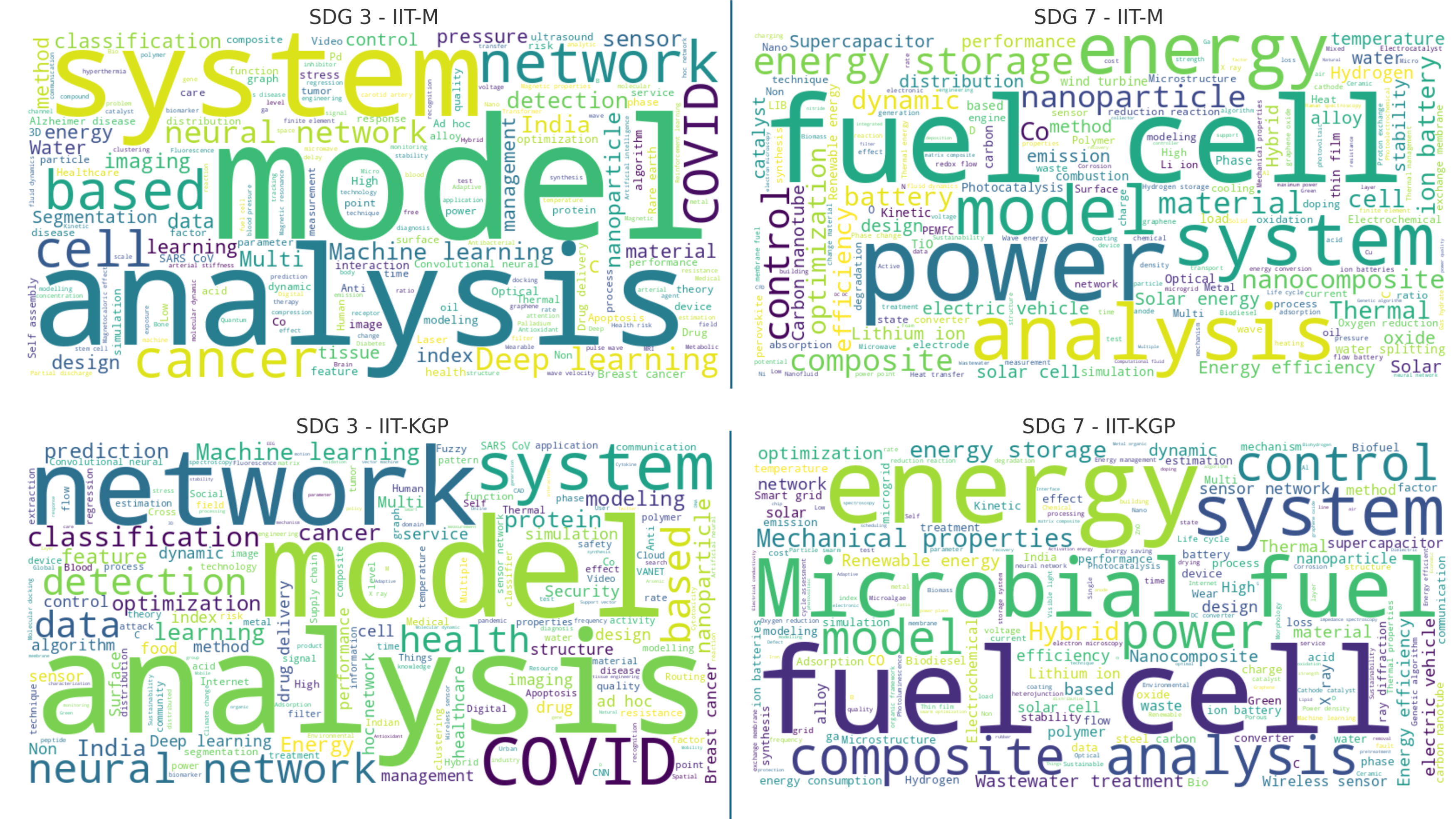} 
\includegraphics[width=0.85\linewidth]{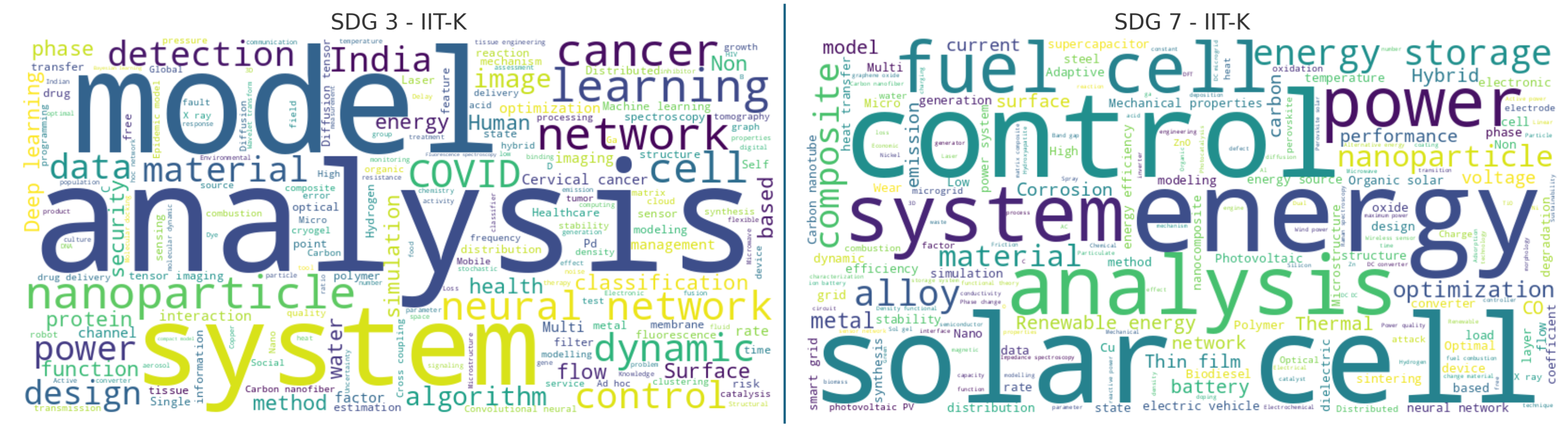} 
\caption{SDG 3 and 7 word cloud.}
\label{fig:SDG_WC}
\end{figure}

Table~\ref{tab:SDG_Goal} provides a detailed breakdown of the contributions of five IITs to the 17 SDGs based on the number and percentage of research papers associated with each goal. IITs collectively focus heavily on SDG 3 (Good Health and Well-being), SDG 7 (Affordable and Clean Energy) and SDG 11 (Sustainable Cities and Communities), as shown in figure~\ref{fig:SDGgraph}. IIT-M leads in health-related research with 18. 63\% of its articles aligned with SDG 3, while IIT-D excels in energy research, contributing 23. 85\% of its articles to SDG 7. Urban sustainability (SDG 11) is a shared focus in all IITs, with contributions averaging around 13\%. In contrast, there is minimal focus on SDG 1 (No Poverty), SDG 5 (Gender Equality), and SDG 16 (Peace, Justice and Strong Institutions), with each receiving less than 1\% contribution in all IITs. IIT-D shows strength in responsible consumption (SDG 12), IIT-KGP leads in water-related research (SDGs 6 and 14), and IIT-K stands out in food security (SDG 2). Despite significant contributions to key SDGs, IITs have opportunities to expand their focus on social and institutional goals such as eradicating poverty, gender equality, and building peace. The main themes reported by ecah IIT in SDG 3 and SDg 7 are shown in figure~\ref{fig:SDG_WC}.

\begin{table}[!h]
\centering
\caption{Contribution to SDG goals by five IITs.}
\resizebox{\textwidth}{!}{ 
\begin{tabular}{|c|cc|cc|cc|cc|cc|}
\hline & \multicolumn{2}{c|}{\textbf{IIT-B}}                                 & \multicolumn{2}{c|}{\textbf{IIT-D}}                                 & \multicolumn{2}{c|}{\textbf{IIT-M}}                                 & \multicolumn{2}{c|}{\textbf{IIT-KGP}}                               & \multicolumn{2}{c|}{\textbf{IIT-K}}                                 \\ \cline{2-11} 
\multirow{-2}{*}{\textbf{\begin{tabular}[c]{@{}c@{}}SDG \\ Goal\end{tabular}}} & \multicolumn{1}{c|}{\textbf{Count}} & \textbf{\%}                   & \multicolumn{1}{c|}{\textbf{Count}} & \textbf{\%}                   & \multicolumn{1}{c|}{\textbf{Count}} & \textbf{\%}                   & \multicolumn{1}{c|}{\textbf{Count}} & \textbf{\%}                   & \multicolumn{1}{c|}{\textbf{Count}} & \textbf{\%}                   \\ \hline
1                                                                              & \multicolumn{1}{c|}{166}            & \cellcolor[HTML]{FBFCFE}0.64  & \multicolumn{1}{c|}{186}            & \cellcolor[HTML]{FBFCFE}0.55  & \multicolumn{1}{c|}{141}            & \cellcolor[HTML]{FBFCFE}0.59  & \multicolumn{1}{c|}{183}            & \cellcolor[HTML]{FBFCFE}0.56  & \multicolumn{1}{c|}{73}             & \cellcolor[HTML]{FCFCFF}0.45  \\ \hline
2                                                                              & \multicolumn{1}{c|}{3135}           & \cellcolor[HTML]{B0DEBE}12.05 & \multicolumn{1}{c|}{3339}           & \cellcolor[HTML]{BFE4CA}9.8   & \multicolumn{1}{c|}{2606}           & \cellcolor[HTML]{B8E1C4}10.9  & \multicolumn{1}{c|}{3707}           & \cellcolor[HTML]{B5E0C2}11.27 & \multicolumn{1}{c|}{2122}           & \cellcolor[HTML]{AADBB8}13.1  \\ \hline
3                                                                              & \multicolumn{1}{c|}{4601}           & \cellcolor[HTML]{8CCF9E}17.68 & \multicolumn{1}{c|}{5240}           & \cellcolor[HTML]{9BD5AB}15.37 & \multicolumn{1}{c|}{4453}           & \cellcolor[HTML]{86CC99}18.63 & \multicolumn{1}{c|}{5327}           & \cellcolor[HTML]{95D3A7}16.19 & \multicolumn{1}{c|}{2985}           & \cellcolor[HTML]{87CD9A}18.42 \\ \hline
4                                                                              & \multicolumn{1}{c|}{589}            & \cellcolor[HTML]{F0F8F5}2.26  & \multicolumn{1}{c|}{463}            & \cellcolor[HTML]{F6FAFA}1.36  & \multicolumn{1}{c|}{441}            & \cellcolor[HTML]{F3F9F7}1.85  & \multicolumn{1}{c|}{587}            & \cellcolor[HTML]{F3F9F8}1.78  & \multicolumn{1}{c|}{303}            & \cellcolor[HTML]{F3F9F7}1.87  \\ \hline
5                                                                              & \multicolumn{1}{c|}{108}            & \cellcolor[HTML]{FCFCFF}0.42  & \multicolumn{1}{c|}{155}            & \cellcolor[HTML]{FCFCFF}0.45  & \multicolumn{1}{c|}{165}            & \cellcolor[HTML]{FAFCFE}0.69  & \multicolumn{1}{c|}{196}            & \cellcolor[HTML]{FBFCFE}0.6   & \multicolumn{1}{c|}{116}            & \cellcolor[HTML]{FAFCFE}0.72  \\ \hline
6                                                                              & \multicolumn{1}{c|}{1567}           & \cellcolor[HTML]{D8EEE0}6.02  & \multicolumn{1}{c|}{1763}           & \cellcolor[HTML]{DDF0E5}5.17  & \multicolumn{1}{c|}{1377}           & \cellcolor[HTML]{D9EEE1}5.76  & \multicolumn{1}{c|}{2398}           & \cellcolor[HTML]{CFEAD9}7.29  & \multicolumn{1}{c|}{903}            & \cellcolor[HTML]{DBEFE2}5.57  \\ \hline
7                                                                              & \multicolumn{1}{c|}{4474}           & \cellcolor[HTML]{8FD0A1}17.19 & \multicolumn{1}{c|}{8127}           & \cellcolor[HTML]{63BE7B}23.85 & \multicolumn{1}{c|}{4339}           & \cellcolor[HTML]{89CE9C}18.15 & \multicolumn{1}{c|}{5029}           & \cellcolor[HTML]{9BD5AC}15.29 & \multicolumn{1}{c|}{2824}           & \cellcolor[HTML]{8DCFA0}17.43 \\ \hline
8                                                                              & \multicolumn{1}{c|}{1197}           & \cellcolor[HTML]{E1F1E8}4.6   & \multicolumn{1}{c|}{1507}           & \cellcolor[HTML]{E2F2E9}4.42  & \multicolumn{1}{c|}{1140}           & \cellcolor[HTML]{E0F1E7}4.77  & \multicolumn{1}{c|}{1481}           & \cellcolor[HTML]{E2F2E8}4.5   & \multicolumn{1}{c|}{770}            & \cellcolor[HTML]{E0F1E7}4.75  \\ \hline
9                                                                              & \multicolumn{1}{c|}{503}            & \cellcolor[HTML]{F2F8F7}1.93  & \multicolumn{1}{c|}{528}            & \cellcolor[HTML]{F5F9F9}1.55  & \multicolumn{1}{c|}{413}            & \cellcolor[HTML]{F4F9F8}1.73  & \multicolumn{1}{c|}{530}            & \cellcolor[HTML]{F4F9F9}1.61  & \multicolumn{1}{c|}{300}            & \cellcolor[HTML]{F3F9F7}1.85  \\ \hline
10                                                                             & \multicolumn{1}{c|}{291}            & \cellcolor[HTML]{F8FBFB}1.12  & \multicolumn{1}{c|}{440}            & \cellcolor[HTML]{F7FAFA}1.29  & \multicolumn{1}{c|}{352}            & \cellcolor[HTML]{F5FAF9}1.47  & \multicolumn{1}{c|}{443}            & \cellcolor[HTML]{F6FAFA}1.35  & \multicolumn{1}{c|}{226}            & \cellcolor[HTML]{F6FAFA}1.39  \\ \hline
11                                                                             & \multicolumn{1}{c|}{3251}           & \cellcolor[HTML]{AEDCBB}12.49 & \multicolumn{1}{c|}{4530}           & \cellcolor[HTML]{A8DAB7}13.29 & \multicolumn{1}{c|}{3137}           & \cellcolor[HTML]{A9DBB8}13.12 & \multicolumn{1}{c|}{4309}           & \cellcolor[HTML]{AADBB8}13.1  & \multicolumn{1}{c|}{2156}           & \cellcolor[HTML]{A8DAB7}13.31 \\ \hline
12                                                                             & \multicolumn{1}{c|}{1966}           & \cellcolor[HTML]{CEEAD7}7.56  & \multicolumn{1}{c|}{3391}           & \cellcolor[HTML]{BEE3CA}9.95  & \multicolumn{1}{c|}{1968}           & \cellcolor[HTML]{C9E8D3}8.23  & \multicolumn{1}{c|}{2857}           & \cellcolor[HTML]{C6E7D1}8.68  & \multicolumn{1}{c|}{1252}           & \cellcolor[HTML]{CDE9D6}7.73  \\ \hline
13                                                                             & \multicolumn{1}{c|}{1300}           & \cellcolor[HTML]{DEF0E5}5     & \multicolumn{1}{c|}{1498}           & \cellcolor[HTML]{E2F2E9}4.4   & \multicolumn{1}{c|}{924}            & \cellcolor[HTML]{E6F3EC}3.87  & \multicolumn{1}{c|}{1568}           & \cellcolor[HTML]{E0F1E7}4.77  & \multicolumn{1}{c|}{554}            & \cellcolor[HTML]{E9F4EE}3.42  \\ \hline
14                                                                             & \multicolumn{1}{c|}{1167}           & \cellcolor[HTML]{E2F2E8}4.49  & \multicolumn{1}{c|}{1085}           & \cellcolor[HTML]{EAF5F0}3.18  & \multicolumn{1}{c|}{1469}           & \cellcolor[HTML]{D7EDDF}6.15  & \multicolumn{1}{c|}{1870}           & \cellcolor[HTML]{DAEEE2}5.68  & \multicolumn{1}{c|}{784}            & \cellcolor[HTML]{DFF1E6}4.84  \\ \hline
15                                                                             & \multicolumn{1}{c|}{1338}           & \cellcolor[HTML]{DDF0E5}5.14  & \multicolumn{1}{c|}{1235}           & \cellcolor[HTML]{E7F4ED}3.62  & \multicolumn{1}{c|}{686}            & \cellcolor[HTML]{ECF6F1}2.87  & \multicolumn{1}{c|}{1805}           & \cellcolor[HTML]{DBEFE3}5.49  & \multicolumn{1}{c|}{594}            & \cellcolor[HTML]{E7F4ED}3.67  \\ \hline
16                                                                             & \multicolumn{1}{c|}{95}             & \cellcolor[HTML]{FCFCFF}0.37  & \multicolumn{1}{c|}{190}            & \cellcolor[HTML]{FBFCFE}0.56  & \multicolumn{1}{c|}{105}            & \cellcolor[HTML]{FCFCFF}0.44  & \multicolumn{1}{c|}{194}            & \cellcolor[HTML]{FBFCFE}0.59  & \multicolumn{1}{c|}{107}            & \cellcolor[HTML]{FBFCFE}0.66  \\ \hline
17                                                                             & \multicolumn{1}{c|}{272}            & \cellcolor[HTML]{F8FBFC}1.05  & \multicolumn{1}{c|}{405}            & \cellcolor[HTML]{F7FAFB}1.19  & \multicolumn{1}{c|}{185}            & \cellcolor[HTML]{FAFBFD}0.77  & \multicolumn{1}{c|}{412}            & \cellcolor[HTML]{F7FAFB}1.25  & \multicolumn{1}{c|}{132}            & \cellcolor[HTML]{FAFBFD}0.81  \\ \hline
\end{tabular}
}
\label{tab:SDG_Goal}
\end{table}

\subsubsection{Country Collaboration: SDG 3 and SDG 7}

\begin{figure}[!h]
    \centering
\includegraphics[width=\linewidth]{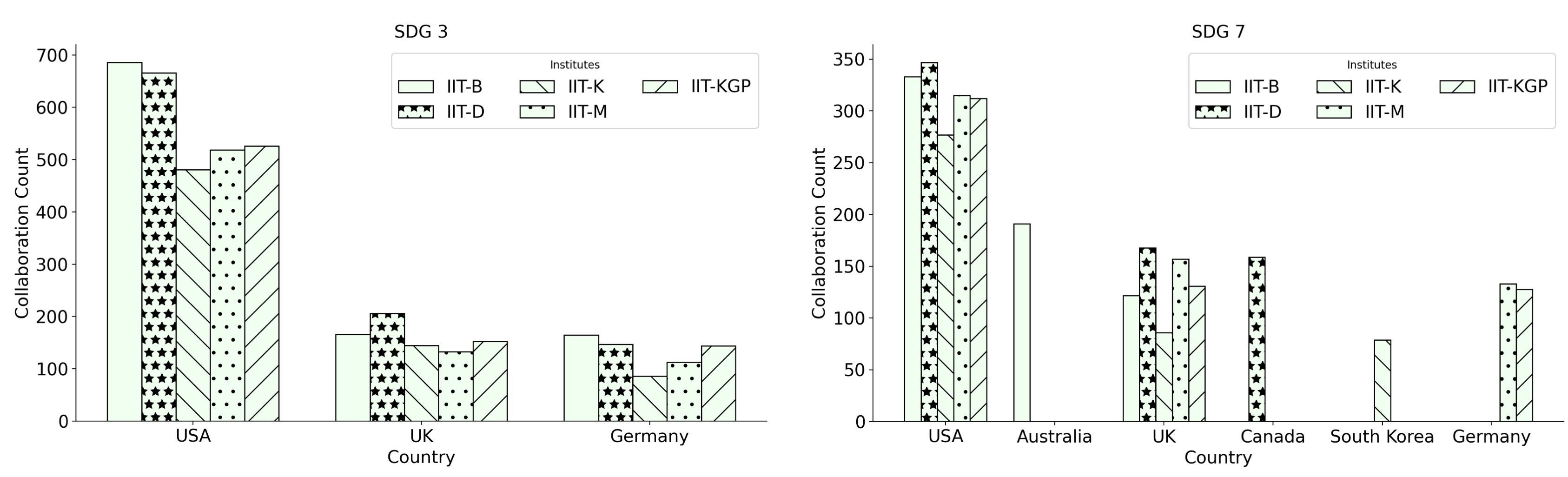} 
\caption{Country Collaboration: SDG 3 and SDG 7.}
\label{fig:SDG-3-7-Colab}
\end{figure}

Figure~\ref{fig:SDG-3-7-Colab} represents the collaboration counts of five IITs with the top three countries collaborators for two SDGs: SDG 3 (Good Health and Well-being) and SDG 7 (Affordable and Clean Energy). In SDG 3 (Figure~\ref{fig:SDG-3-7-Colab}(left))shows the highest collaboration counts in all IITs, with IIT-D and IIT-B having particularly strong contributions. UK emerges as the second most significant collaborator for IIT-B, IIT-D, and IIT-K, while Germany is the second most significant collaborator for IIT-M and IIT-KGP. For SDG 7 (Figure~\ref{fig:SDG-3-7-Colab}(right)), the USA continues its dominance as the main collaborator of all IITs, with IIT-B and IIT-D making substantial contributions. The second and third positions vary by IIT, reflecting differences in regional engagement. This diversity underscores the global nature of SDG-aligned research conducted by the IITs. Across both SDGs, the USA holds the first position for all IITs, underlining its role as the most significant research partner.

\section{Conclusion}

The Indian Institutes of Technology (IITs) have become the pillars of India's research ecosystem, consistently making significant contributions to global science and technology. Their exponential growth in research productivity over the decades reflects the increasing emphasis on quality, funding, and strategic collaborations. IIT-Kanpur, IIT-Kharagpur, and IIT-Delhi have established themselves as leaders in both productivity and impact, while IIT-Bombay and IIT-Madras continue to showcase their dominance in niche areas such as optical materials and fluid dynamics.

Collaborations, particularly with western countries and Asian partners, have significantly increased the global reach and impact of IIT research on the citation. Interdisciplinary approaches integrating computational methods with traditional engineering and sciences highlight the innovative spirit of these institutions. The USA, UK, and Germany are the most frequent collaborators with IITs, underlining their global research connections. Regional collaborators such as Canada, Australia, and South Korea also play significant roles, albeit with specific IITs and research domains. This reflects the global and goal-specific nature of IITs' international research collaborations.

Collaboration among IITs shows a strong preference for partnerships between older and more established institutions (Bombay, Delhi, Kanpur, Kharagpur, and Madras), likely due to historical ties, increased research capacity, and resource availability. To foster stronger collaboration with newer IITs, older IITs can initiate joint research projects focusing on shared interests such as clean energy, healthcare, and emerging technologies. Resource sharing, including access to advanced laboratories and computational tools, can enable newer IITs to engage in high-impact research. Faculty exchange programs and mentoring initiatives can strengthen the research capabilities of newer IITs, while joint centers of excellence can unite the strengths of both groups to address national and global challenges. 

 IITs collectively contribute significantly to the global goals of the SDGs, particularly in health, energy, and urban development. However, there is a noticeable lack of focus on social and institutional goals such as poverty eradication, gender equality, and peacebuilding, which presents opportunities for more diversified research in the future.
\subsection{Limitations}
Although the study provides valuable information on research trends and the impact of IITs, certain limitations must be recognized to contextualize its findings. First, the study relies on Scopus-indexed publications, potentially overlooking impactful research published in other indexed databases like WoS, Dimension, and nonindexed or regional journals. Second, only the top five IITs are analyzed, which may not capture the contributions of newer IITs or specific regional variations. Third, the study primarily emphasizes international collaborations and inter-IIT collaboration and may underrepresent domestic or industry-academic partnerships.

\section*{Acknowledgment}
Dr. Kiran gratefully acknowledges the invaluable learning support provided by the Centre for Teaching, Learning, and Development at BML Munjal University. Special thanks are extended to the Research and Development Cell for their financial support through the seed grant (No: BMU/RDC/SG/2024-06), which made this research possible.

\section*{Conflict of interest}

All authors declare no conflict of interest.

\section*{Data availability}
The data utilized in this study is accessible for reproducibility upon request from the corresponding and principal authors.
\appendix
\renewcommand{\thetable}{Appendix \arabic{table}}
\section*{Appendix}
\begin{table}[!h]
\centering
\caption{List of all ITTS with Scopus affiliation code and abbreviated name.}
\begin{tabular}{|l|l|l|}
\hline
\textbf{Code} & \textbf{Abbreviation} & \textbf{IITs}                                                       \\ \hline
60032730      & \textbf{IIT D}        & Indian Institute of Technology, Delhi                               \\ \hline
60014153      & \textbf{IIT B}        & Indian Institute of Technology, Bombay                              \\ \hline
60004750      & \textbf{IIT KGP}      & Indian Institute of Technology,  Kharagpur                          \\ \hline
60021988      & \textbf{IIT K}        & Indian Institute of Technology,  Kanpur                             \\ \hline
60025757      & \textbf{IIT M}        & Indian Institude of Technology,  Madras                             \\ \hline
60031818      & \textbf{IIT R}        & Indian Institude of Technology,  Roorkee                            \\ \hline
60010126      & \textbf{IIT G}        & Indian Institude of Technology, Guwahati                            \\ \hline
60104350      & \textbf{IIT I}        & Indian Institude of Technology,  Indore                             \\ \hline
60103917      & \textbf{IIT H}        & Indian Institude of Technology, Hyderabad                           \\ \hline
60104339      & \textbf{IIT BBS}      & Indian Institude of Technology, Bhubaneshwar                        \\ \hline
60104342      & \textbf{IIT P}        & Indian Institude of Technology,  Patna                              \\ \hline
60104343      & \textbf{IIT J}        & Indian Institude of Technology,  Jodhpur                            \\ \hline
60104341      & \textbf{IIT GN}       & Indian Institude of Technology,  Gandhinagar                        \\ \hline
60104340      & \textbf{IIT MANDI}    & Indian Institude of Technology,  Mandi                              \\ \hline
60103918      & \textbf{IIT RPR}      & Indian Institude of Technology, Ropar                               \\ \hline
60019106      & \textbf{IIT V}        & Indian Institute of Technology (Banaras Hindu University), Varanasi \\ \hline
60109702      & \textbf{IIT JU}       & Indian Institute of Technology, Jammu                               \\ \hline
60109689      & \textbf{IIT PD}       & Indian Institute of Technology, Palakkad                            \\ \hline
60109690      & \textbf{IIT T}        & Indian Institute of Technology, Tirupati                            \\ \hline
60114558      & \textbf{IIT GA}       & Indian Institute of Technology, Goa                                 \\ \hline
60114557      & \textbf{IIT BI}       & Indian Institute of Technology, Bhilai                              \\ \hline
60114348      & \textbf{IIT DHD}      & Indian Institute of Technology, Dharwad                             \\ \hline
60008898      & \textbf{IIT DBD}      & Indian Institute of Technology (Indian School of Mines),   Dhanbad  \\ \hline
\end{tabular}
\label{Table:IITName}
\end{table}


\begin{table}[!h]
\centering
\caption{Lis of prominent India funding bodies and abbreviation and full name.}
\begin{tabular}{|l|l|}
\hline
\textbf{Funding Body} & \textbf{Agency Name}                                                                                      \\ \hline
\textbf{DST India}    & Department of Science and Technology, Ministry of Science and Technology                                  \\ \hline
\textbf{IIT}          & Indian Institute of Technology                                                                            \\ \hline
\textbf{CSIR}         & Council of Scientific and Industrial Research                                                             \\ \hline
\textbf{NSF}          & National Science Foundation                                                                               \\ \hline
\textbf{SERB}         & \cellcolor[HTML]{FFFFFF}Science and Engineering Research Board                                            \\ \hline
\textbf{DBT}          & \cellcolor[HTML]{FFFFFF}Department of Biotechnology, Ministry of Science and Technology                   \\ \hline
\textbf{DST Kerala}   & \cellcolor[HTML]{FFFFFF}{\color[HTML]{1F1F1F} Department of Science and Technology, Government of Kerala} \\ \hline
\textbf{UGC}          & \cellcolor[HTML]{FFFFFF}{\color[HTML]{1F1F1F} University Grants Commission}                               \\ \hline
\textbf{MoE}          & \cellcolor[HTML]{FFFFFF}{\color[HTML]{1F1F1F} Ministry of Education}                                      \\ \hline
\textbf{MHRD}         & \cellcolor[HTML]{FFFFFF}{\color[HTML]{1F1F1F} Ministry of Human Resource Development}                     \\ \hline
\textbf{MNRE}         & \cellcolor[HTML]{FFFFFF}{\color[HTML]{1F1F1F} Ministry of New and Renewable Energy}                       \\ \hline
\textbf{ISRO}         & Indian Space Research Organisation                                                                        \\ \hline
\textbf{DAE}          & \cellcolor[HTML]{FFFFFF}{\color[HTML]{1F1F1F} Department of Atomic Energy, Government of}                 \\ \hline
\textbf{MEITy}        & \cellcolor[HTML]{FFFFFF}Ministry of Electronics and Information Technology                                \\ \hline
\textbf{DRDO}         & \cellcolor[HTML]{FFFFFF}Defence Research and Development Organisation                                     \\ \hline
\end{tabular}
\label{table:appendix}
\end{table}
\bibliographystyle{cas-model2-names}
\bibliography{cas-refs}

\end{document}